\documentclass[english,11pt,a4paper]{article}
\usepackage{jheppub} 
\usepackage[T1]{fontenc}
\usepackage[font=small,labelfont=bf]{caption}
\usepackage{enumitem}

\renewcommand{\eqref}[1]{Eq.~(\ref{#1})}

\newcommand{\secref}[1]{Sec.~\ref{sec:#1}}
\newcommand{\secsref}[2]{Secs.~\ref{sec:#1} and \ref{sec:#2}}

\newcommand{\appref}[1]{Appendix~\ref{sec:#1}}
\newcommand{\figref}[1]{Fig.~\ref{fig:#1}}

\newcommand{\Figref}[1]{Figure~\ref{fig:#1}}
\newcommand{\tableref}[1]{Table~\ref{tab:#1}}

\newcommand{\beqa}{\begin{align}} 
\newcommand{\eeqa}{\end{align}}

\newcommand{\beq}{\begin{equation}}
 \newcommand{\eeq}{\end{equation}}
\newcommand{\bpm}{\begin{pmatrix}}
 \newcommand{\epm}{\end{pmatrix}}
\newcommand{\nn}{\nonumber}

\def\kahler{K\"ahler }
\def\softsusy{\texttt{SOFTSUSY}}
\def\feynhiggs{\texttt{FeynHiggs}}

\newcommand{\Yud}{{Y_u^{\rm diag}}}
\newcommand{\Ydd}{{Y_d^{\rm diag}}}
\newcommand{\yud}{{y_u^{\rm diag}}}
\newcommand{\Vckm}{{V_{\rm CKM}}}
\newcommand{\Vul}{{V_{uL}}}
\newcommand{\Vur}{{V_{uR}}}
\newcommand{\Vdl}{{V_{dL}}}
\newcommand{\Vdr}{{V_{dR}}}
\newcommand{\Wul}{{W_{uL}}}
\newcommand{\Wur}{{W_{uR}}}

\newcommand{\vev}[1]{\langle #1\rangle}
\newcommand{\Tr}{\mathop{\mathrm{Tr}}}
\newcommand{\Order}{\mathop{O}}
\newcommand{\dd}{\mathrm{d}}

\def\GeV{\text{ GeV}}
\def\TeV{\text{ TeV}}
\title{LHC Benchmarks from Flavored Gauge Mediation}

\author{N. Ierushalmi,}
\author{S. Iwamoto,}
\author{G. Lee,}
\author{V. Nepomnyashy}
\author{and Y. Shadmi}
\affiliation{Physics Department, Technion---Israel Institute of Technology,\\ Haifa 32000, Israel}

\emailAdd{nivieru@technion.ac.il}
\emailAdd{sho@physics.technion.ac.il}
\emailAdd{leeg@physics.technion.ac.il}
\emailAdd{vera.nepomnyashy@gmail.com}
\emailAdd{yshadmi@physics.technion.ac.il}

\abstract{%
We present benchmark points for LHC searches from flavored gauge
mediation models, in which messenger--matter couplings give
flavor-dependent squark masses. Our examples include spectra in which a
single squark---stop, scharm,
or sup---is much lighter than all other colored superpartners, motivating
improved quark flavor tagging at the LHC. Many examples feature flavor
mixing; in particular, large stop--scharm mixing is possible. The correct
Higgs mass is obtained in some examples by virtue of the large stop A-term.
We also revisit the general flavor and CP structure of the models.
Even though the A-terms can be substantial,
their contributions to EDM's are very suppressed,
because of the particular dependence of the A-terms on the messenger coupling.
This holds regardless of the messenger-coupling texture.
More generally, the special structure of the soft terms often
leads to stronger suppression of flavor- and CP-violating processes, 
compared to naive estimates.
}

\begin{document}
\maketitle


 \section{Introduction}
 \label{sec:intro}
Flavored gauge mediation (FGM) models~\cite{Shadmi:2011hs}
extend minimal gauge mediation~\cite{Dine:1994vc,Dine:1995ag}
 by introducing superpotential couplings 
between the messenger and matter superfields~\cite{Dine:1996xk,Chacko:2001km,Joaquim:2006uz,Joaquim:2006mn,Brignole:2010nh}.
Generically, such couplings lead to flavor-dependent sfermion spectra.
They also generate A-terms at the messenger-scale, allowing for
a 125~GeV Higgs with relatively light squarks~\cite{Evans:2011bea,Evans:2011uq,Evans:2012hg,Kang:2012ra,Evans:2012uf,Craig:2012xp,Albaid:2012qk,Abdullah:2012tq,Byakti:2013ti,Craig:2013wga,Evans:2013kxa,Jelinski:2015voa,Basirnia:2015vga}.
The main observation of~\cite{Shadmi:2011hs} is that any concrete mechanism  
for generating the standard model (SM) Yukawas also governs the textures of the 
messenger--matter couplings, and can therefore give spectra consistent with low-energy bounds.
The messenger coupling is either similar in structure to the corresponding 
SM Yukawa texture, resulting in minimal flavor violation (MFV)-like models
in which only the third-generation sfermion masses are affected (with the possiblility
of large mixings between different generations of sfermions), 
or it has an altogether different structure,
leading to large effects in the first- and second-generation sfermions.
Flavor constraints on MFV-like models were analyzed in detail
in~\cite{Calibbi:2013mka,Calibbi:2014yha}.
Non-MFV models were explored in~\cite{Galon:2013jba},
and shown to give unusual squark spectra, 
featuring for example a single up or charm squark much lighter than the 
remaining squarks, and/or large stop--scharm mixing,
with important implications for LHC supersymmetry searches~\cite{Mahbubani:2012qq,Blanke:2013zxo,Blanke:2015ulx}.
Flavor effects in different extensions of gauge mediated supersymmetry breaking (GMSB) models were also considered
in~\cite{Brummer:2013upa,Jelinski:2014uba,Abel:2014fka,Evans:2015swa}.

Our main goal in this paper is to extend the qualitative analysis of~\cite{Galon:2013jba},
and to provide interesting FGM benchmarks for the LHC run~II.
We derive several representative examples by starting with specific flavor models
at the high scale, and including RGE effects above and below the messenger scale. 
We find that even MFV-like models have a rich structure which allows
for large stop hierarchies and a large stop mixing.
Examples of this type were also discussed recently in~\cite{Backovic:2015rwa}.
Our examples typically feature the gluino and most squarks
at or above 2\,TeV, with one or two squarks at or below 1\,TeV.
Flavor mixing generically occurs in parts of the spectra.
Many of these examples, including all the MFV-like models and 
some of the non-MFV models, also give the correct Higgs mass.

Regardless of the texture of the messenger--matter coupling,
the fact that the soft terms are all determined by a single new coupling
matrix (in addition to the gauge and SM Yukawa couplings) leads to
several remarkable features in the flavor and CP structure of the models.
First, the A-terms do not generate electric dipole moments (EDMs) at leading order
because they only depend on the absolute value squared of the messenger coupling~\cite{Calibbi:2014yha}.
Thus, the lack of a supersymmetric CP problem, one of the nicest features
of GMSB, is not spoiled in these models.
This holds more generally for other extensions of GMSB that introduce a
single dominant messenger--SM--SM coupling.
Second, this special structure also results in extra protection 
against flavor-non-diagonal CP violation.
In particular, if there is a single non-zero entry in the messenger coupling 
matrix, it does not introduce any new CP-violating phase,
and the only CP violation originates from the SM Yukawas.
In the LL sector, the only source of CP violation is therefore the CKM matrix,
and CP- and flavor-violating processes in the first and second generations are very small.
Finally, in models with a vanishing (or small) down-type messenger coupling,
the new contributions to the R-down squark masses involve two powers of 
the SM down Yukawa, so these masses are barely modified from their GMSB values. 
The constraints on the combination of LL and RR flavor parameters in 
the down sector, which are particularly stringent, 
are therefore automatically met.

An important ingredient in FGM models
is a mechanism for generating the SM Yukawas.
As we review in \secref{FGM}, since the SM Yukawas have a hierarchical structure,
the messenger--matter couplings are generically hierarchical too, 
with at most a single entry being $\mathcal{O}(1)$.
The existence and location of this entry in the messenger--matter coupling matrix
depends on the details of the model. 
As a concrete example, we assume that fermion masses are governed by
a horizontal U(1) symmetry,
 broken by a spurion $\lambda\sim0.2$ of charge $-1$.
 The fermion masses are then proportional to different powers of the
 spurion $\lambda$,
with the powers determined by the flavor charges of the matter and Higgs
fields.
Once the messenger flavor charge is specified, 
the messenger--matter coupling is completely determined too.
There are then three possibilities:
\begin{enumerate}
\item The messenger charge is the same as the Higgs charge.
  Then the texture of the messenger-coupling matrix $y_U$ is identical to
  the Yukawa texture $Y_U$,
and only its 3--3 entry is order one.
This leads to the MFV-like models of~\cite{Shadmi:2011hs,Abdullah:2012tq}.
Since the only large effects are in the third generation,
the models are consistent with low-energy constraints.
\item The messenger charge is larger than the Higgs charge.
  Then the messenger couplings have larger suppressions compared to the Yukawa
  couplings, $y_{Uij}<Y_{Uij}$.
Since the messenger couplings enter squared in the soft terms, 
their effects on the masses are very small, and again, 
the models are consistent with flavor bounds. 
Note that, even though the mass splittings are small in this case,
flavor mixing can be large. The reason is that the dominant contribution to sfermion
masses is the gauge-mediated contribution, which is proportional to the identity.
The mixing is therefore solely determined by the corrections to the GMSB
contribution~\cite{Feng:2007ke,Shadmi:2011hs}.
\item The messenger charge is smaller than the Higgs charge.
Then one of the messenger couplings involving the first or second generation can be
$\mathcal{O}(1)$,
giving a large mass splitting to the squarks of the respective generation. 
This requires precise down alignment in order to satisfy flavor constraints.
\end{enumerate}
In the examples below, we choose different messenger charges,
thus sampling different entries of the messenger coupling matrix, and covering
most of the FGM parameter space.
In each case, one can get a viable spectrum consistent with flavor constraints.

Note that, by construction, all the models exhibit some degree of supersymmetric alignment~\cite{Abdullah:2012tq}:  
the sfermion mass matrices and the SM Yukawa matrices are simultaneously diagonal
up to corrections of order $\lambda\sim0.2$. This alignment is sufficient to
satisfy flavor constraints in models of types (1) and (2). 
The non-MFV models further require zero 1--2 L-down mixing.
We refer to this alignment as \textit{supersymmetric alignment}~\cite{Abdullah:2012tq,Galon:2013jba} 
because the supersymmetric messenger couplings, 
rather than the supersymmetry-breaking parameters~\cite{Nir:1993mx},
are controlled by the flavor symmetry.
As a result, the flavor symmetry is manifest at the (possibly) low messenger scale.
A generic feature of these models is O($\lambda$) mixing in the first and second generation
L-squarks, either in the up sector only~\cite{Ghosh:2015nui}, or, in models of type (1)
and (2), in both the up and down sector.

We present several examples with non-trivial squark spectra
for $\tan\beta = 10$ and one or two messenger pairs. 
Our benchmark points (BPs) have negligible messenger--lepton couplings, so the slepton masses 
remain degenerate as in GMSB.
The NLSP in these examples is either a bino or an R-stau.
The latter possibility can be realized even in models with a single messenger
pair, since the splitting in the squark masses due to the new messenger 
coupling feeds into the slepton masses through the hypercharge RGE 
contribution~\cite{Evans:2011bea,Abdullah:2012tq}.
As usual in GMSB models, the NLSP lifetime depends on the gravitino mass,
which in turn depends on the underlying supersymmetry-breaking scale.
In the examples we show, this uncertainty translates into a wide range
of bino or R-slepton lifetimes, underscoring the importance of
 searches for intermediate non-prompt decays (see also~\cite{Evans:2016zau} for a recent
discussion).

Since our main focus is squark flavor, we also show examples in which the Higgs
has a mass below 125~GeV, requiring an additional modification of the model to 
raise it, such as the addition of an NMSSM singlet.
Such modifications will have little effect on the sfermion flavor structure. 
The large Higgs mass in our models is driven by the large stop LR mixing,
and in some examples,  by the large stop masses.
To get a handle on the uncertainty in the calculated Higgs mass, 
we computed it
using both \softsusy~\cite{Allanach:2001kg} and \feynhiggs~\cite{Heinemeyer:1998yj,
Heinemeyer:1998np,Degrassi:2002fi,Frank:2006yh,Hahn:2013ria}.
We collect the comparisons between the two codes in Appendix~\ref{sec:feynhiggs}.
Furthermore, the large stop--scharm mixing and the large 3--2 entry of the
up A-term featured in some of the models may modify the Higgs mass
by a few GeV~\cite{Cao:2006xb,Kowalska:2014opa,Brignole:2015kva,Goodsell:2015yca}. 
Thus, the large Higgs mass values we obtain in the models should mainly
be viewed as an indication that the models can accommodate a correct Higgs mass.

This paper is organized as follows.
In~\secref{FGM}  we discuss the general flavor
and CP structure of the models. In~\secref{MFV} we present several examples
of MFV-like models, and the resulting low energy spectra.
Non-MFV examples are presented in~\secref{NonMFV}.
We conclude with some remarks in~\secref{concl}.
Appendix~\ref{sec:fgmapp} reviews some basics of  FGM models, including possible symmetries
yielding the FGM superpotential, and the issue of
superpotential and K\"ahler mixing. 
Our conventions for the soft terms, as well as the full expressions for the soft terms
in the presence of up-type messenger couplings, are collected in
Appendix~\ref{sec:soft}.
In Appendix~\ref{sec:feynhiggs} we compare the values of the Higgs mass as computed
by \softsusy~and \feynhiggs. The values of the flavor-violating $\delta$'s
for our different examples are collected in Appendix~\ref{sec:app_deltas}.


\section{General structure: flavor and CP}
 \label{sec:FGM}
We consider models with the superpotential~\cite{Chacko:2001km,Shadmi:2011hs}
\begin{equation}
\begin{split}
   W = \sum_{I=1}^{N_5}X(T_I \bar T_I + D_I \bar D_I )
 &+ \bar D_1\, q\, y_U\, u^c + D_2\, q\, y_D\, d^c + D_2\, l\, y_L\, e^c\\
 &+ H_U\, q\, Y_U\, u^c + H_D\, q\, Y_D\, d^c + H_D\, l\, Y_L\, e^c\,.
\end{split}\label{superpot}
\end{equation}
Here $X$ parameterizes the supersymmetry breaking with $\vev{X}=M+F\theta^{2}$,
($T_I$, $\bar T_I$) and ($D_I$, $\bar D_I$) are respectively
SU(3) triplet and SU(2) doublet messenger pairs, and $N_5$ is the number of messenger pairs.
$Y_{U,D,L}$ are the SM Yukawa coupling matrices, and $y_{U,D,L}$
are the analogous 3$\times3$ flavor-space matrices of messenger--matter couplings;
hereafter we use instead the complex conjugates of these couplings,
\begin{equation}
 Y_{u,d,l} = (Y_{U,D,L})^*,\qquad
 y_{u,d,l} = (y_{U,D,L})^*,
\end{equation}
for easier comparison with the SM flavor constraints.
Our main focus below is on examples with $y_d=y_l=0$, but in this section we discuss some general features of the models with both $y_u$ and $y_d$ present.
The form of the superpotential~(\ref{superpot}) can be enforced 
by various choices of global
symmetries, as we review in Appendix~\ref{sec:fgmapp}.
The expressions for the soft terms, for $y_d=y_l=0$, are collected in Appendix~\ref{sec:soft}.
For simplicity, we denote $\bar D\equiv\bar D_1$.

The new couplings generate one-loop A-terms at the messenger scale as well as new
contributions to the sfermion soft masses.
Unlike the GMSB contributions, these are generically flavor dependent, and contain new
sources of CP violation.
In~\tableref{boundsMM}, 
we summarize the mass insertion approximation (MIA) estimates for the
most stringently constrainted squark flavor parameters 
from~\cite{Isidori:2010kg,Galon:2013jba}.
Working in terms of the physical masses and mixings, we define~\cite{Raz:2002zx}
 \beq\label{defdel}
 \bigl( \delta_{ij}^q \bigr) _{MM} = \frac{\Delta \tilde m^2_{ji}}{\tilde m^2_q} \bigl( K_M^q \bigr) _{ij} \bigl( K_M^q \bigr) ^* _{jj} \,,
 \eeq
where $\Delta \tilde m^2_{ji} = m^2_{\tilde q_j} - m^2_{\tilde q_i}$ is the
squared-mass difference of the relevant squarks, 
$\tilde m^2_q = \frac{1}{3} \sum_{\alpha=1}^3 m^2_{\tilde q_\alpha}$ is the average squark squared-mass,
and $(K_M^q)_{ij}$ is the mixing appearing in the quark--squark--gluino coupling.
In the models considered below, there is a single dominant source of 
flavor violation, so the MIA gives a reasonable estimate.
As is well known, this approximation breaks down with $\mathcal{O}(1)$ mass differences
(see for example~\cite{Calibbi:2015kja,Ghosh:2015nui} for recent analyses).
In some of the examples below, the relative mass splitting is very small,
so the MIA is reliable. In the others, with $\mathcal{O}(1)$ mass differences, the mixings are very small,
so that even though the MIA does not give a good estimate, flavor constraints are
satisfied thanks to the small mixings.
%
%
 \begin{table}[tp]
 \centering
 \renewcommand{\arraystretch}{1.3}
 \begin{tabular}{|cc|ccc|}
 \hline
 $q$ & $ij$ & $| (\delta^q_{ij})_{MM}|$ & 
 $\sqrt{{\rm Im}[(\delta^q_{ij})_{MM}^2]}$ & 
 $ \sqrt{{\rm Im}[(\delta^q_{ij})_{LL} (\delta^q_{ij})_{RR}]}$ \\
 \hline
 $d$ & 12& $0.07$ & $0.01$ & $0.0005$\\
 $u$ & 12 & $0.1$& $0.05$ & $0.003$ \\
 $d$ & 23 & $0.6$ & $0.2$ & $0.07$\\
 \hline
 \end{tabular}
  \caption{Bounds on $(\delta^q_{ij})_{MM}$ 
 for 1\,TeV common squark and gluino masses.
 For higher squark masses the bounds scale approximately as $\tilde{m}_{q}$, 
with a weaker dependence on the gluino mass.
 }
 \label{tab:boundsMM}
 \end{table}

The flavor textures of the soft terms are not arbitrary in these models.
Rather, they are given by specific combinations of the messenger couplings and the SM Yukawas,
which can be determined by a spurion analysis, treating the matrices
$Y_u$, $y_u$, etc.\ as spurions of the SU(3)$^5$ flavor symmetry~\cite{Shadmi:2011hs}. 
Furthermore, any power of the messenger coupling appearing in the soft terms, 
say $y_u$, must be accompanied by the same power of $y_u^\dagger$~\cite{Calibbi:2014yha}.
To see this, one can invoke a global U(1) under which only the messenger 
field $\bar D_1$ is charged. $y_u$ is the only spurion breaking this U(1).
Since the soft terms are singlets of this symmetry, $y_u$ must appear 
with $y_u^\dagger$.
This special structure ameliorates some of the constraints on the
models, especially in the presence of a single new coupling, as we now discuss.

First, the models do not suffer from the flavor-diagonal supersymmetric CP problem.
Indeed, one of the virtues of GMSB models is that they do not introduce large quark and lepton EDMs.
In minimal GMSB, the A-terms are only generated from the gaugino mass through the RGE,
so there is no relative phase between them.
Naively, this seems to be spoiled in FGM models (and more generally in models with 
messenger--matter or messenger--Higgs couplings),
since the messenger--matter couplings generate A-terms at the messenger scale.
However, to leading order, these A-terms are real~\cite{Calibbi:2014yha}.
For example, consider the $y_u$ contribution to $A_d$ at one-loop order.
It is determined by the spurion analysis described above as
\begin{equation}
 A_d^* \propto y_u y_u^\dagger Y_d.
\end{equation}
To discuss the resulting EDMs, we rotate the squark superfields into the quark mass basis,
in which the SM Yukawa matrices are diagonal.
In this basis, the A-term is given by
\begin{equation}
 \Vdl A_d^* \Vdr^\dagger \propto\Vdl\,y_u y_u^\dagger\Vdl^\dagger\, \Ydd\,,
\end{equation}
where $\Vdl$, $\Vdr$ ($\Vul$, $\Vur$) are the matrices that bidiagonalize the
down and up SM Yukawa matrices,
\begin{equation}
 \Yud = \Vul\, Y_u\, \Vur^\dagger\,,~~\Ydd = \Vdl\, Y_d \,\Vdr^\dagger,
\end{equation}
such that the CKM matrix is given by $\Vckm=\Vul  \Vdl^\dagger$.
Since $\Vdl\,y_u y_u^\dagger\Vdl^\dagger$ is hermitian,
\beq
{(\Vdl A_d^* \Vur^\dagger)}_{11} \propto {(\Vdl y_u y_u^\dagger\Vdl^\dagger)}_{11} {(\Ydd)}_{11}
\eeq
has no complex phase and does not generate an EDM.
Similarly, the $y_u$ contribution to $A_u$ is restricted to be of the form
\beq
 (y_u y_u^\dagger) Y_u\,,~~~~~ Y_u  (y_u^\dagger y_u)\,,
\eeq
and thus $(A_u)_{11}$ evaluated in the fermion mass basis is real.
We can repeat this discussion for the contributions from $y_d$.
The same reasoning also applies to any model with a messenger--SM--SM coupling. 
A non-zero EDM can, in principle, be generated at higher orders in the messenger loops,
coming either from higher-loop corrections to the A-terms, or from multiple
insertions of $\delta_{LR}$, $\delta_{LL}$, $\delta_{RR}$,
but these are very small.\footnote{Many of the possible structures are actually real.
In fact, with $y_d=0$, no imaginary part can be generated if $y_u$ is rank 1.}

We now turn to flavor violation. As noted above, the combination of flavor violation 
in the L- and R-down squarks is tightly constrained by experiment.
These constraints are alleviated in our models for $y_d=0$,
which can be achieved by a choice of global symmetries (see
Appendix~\ref{sec:fgmapp}).
With no down-type messenger couplings, the new contribution to the RR down squark mass 
matrix~\eqref{app_sqmasses}, is suppressed by two powers of the down Yukawa.
In the fermion mass basis, this mass matrix is given by
\begin{equation}
 \Vdr (\delta\tilde m^2_{d_R}) \Vdr^\dagger\,.
 \end{equation}
Thus for example, its 1--2 entry is
\begin{equation}
 \left(\Vdr (\delta\tilde m^2_{d_R}) \Vdr^\dagger\right)_{12}
  \sim \frac{m_d m_s}{\vev{H_D}^2}
(\Vdl\, y_u y_u^\dagger\, \Vdl^\dagger)_{12}\sim  10^{-8}\tan^2\beta
(\Vdl\, y_u y_u^\dagger\, \Vdl^\dagger)_{12}\,,
\end{equation}
where $m_d$ ($m_s$) is the down (strange) quark mass and $\vev{H_D}$ is the vacuum expectation value of $H_D$.
This is negligible even for an arbitrary ${\cal O}(1)$ matrix $y_u$,
so that the R down squarks are nearly degenerate.

The remaining LL and RR mass matrices involve 
various combinations of $y_u$ and the SM Yukawas.
First, consider terms that only involve $y_u$.
In the fermion mass basis, these have the form
\begin{align}
\label{sqmasses}
 \tilde m^2_{d_L} &\supset
(\Vdl\, \Wul^\dagger) \left[ ( \yud \yud^\dagger) + \# {(\yud \yud^\dagger)}^2
\right]
{(\Vdl\, \Wul^\dagger)}^\dagger\,,\\
 \tilde m^2_{U_L} &\supset
(\Vul\Wul^\dagger)\, \left[
(\yud \yud^\dagger) +\#  (\yud \yud^\dagger)^2\,
\right]
 {(\Vul\Wul^\dagger)}^\dagger\\
 \tilde m^2_{U_R} &\supset
 (\Vur\, \Wur^\dagger) \left[
 (\yud^\dagger \yud) +\# (\yud^\dagger \yud)^2
\right]
{(\Vur\Wur^\dagger)}^\dagger\,,
\end{align}
where the $\#$ are real numbers,
and we define
\beq\label{wmix}
\yud \equiv \Wul\, y_u\, \Wur^\dagger
\eeq
with $\yud$ diagonal.
Clearly, flavor mixings are determined by the misalignment between $y_u$ 
and the relevant Yukawa matrix.
The most stringent constraint is on $(\delta^d_{12})_{LL}$.
In the following, we will consider two types of models. 
\begin{itemize}
\item In MFV-like models, $(y_u)_{ij}$ and $(Y_u)_{ij}$ are the same up 
to order-one numbers, for any $i,j$.
Then the only large entry of $y_u$ is the 33 entry, with 
\beq
\yud \yud^\dagger\sim{\rm diag}(0, \lambda^4,1),
\eeq
where $\lambda\sim0.2$.
Also, without accidental cancellations or enhancements, $(\Wul\Vdl^\dagger)_{ij}$
is expected to be the same order as $(\Vckm)_{ij}$. Then
$(\delta^d_{12})_{LL}\sim\lambda^5$, which is at the level of experimental constraints,
and $(\delta^d_{23})_{LL}\sim\lambda^2$, which is below current sensitivity.
The MFV-like models are thus consistent with flavor bounds on the down sector.

As for the up sector, in MFV-like models the 1--2 up block is approximately 
degenerate, and this degeneracy is sufficient for satisfying 
all flavor constraints.
Furthermore, generically in these models  $\Vul\sim\Wul$ and $\Wur\sim \Vur$,
so that
\beq
 (\Vul\Wul^\dagger) \sim \Vul\,,~~~~(\Vur\, \Wur^\dagger)\sim \Vur\,.
\eeq

\item In other models, $y_u$ has $\mathcal{O}(1)$ entries in the first or second row or column.
Then $\yud \yud^\dagger$ has $\mathcal{O}(1)$ entries in its 1--2 block, and 
precise alignment of the down squark and quark matrices is required 
for  ${(\Vdl\, \Wul^\dagger)}_{12}$ to be small.
We therefore construct models in which the down Yukawa is approximately diagonal,
and $y_u y_u^\dagger$ is approximately diagonal as well. Then, $\Vdl$, $\Wul$, and
$\Wur$ are close to the identity matrix, and flavor mixing arises predominantly
from the SM mixing matrices $\Vul\sim \Vckm$ and $\Vur$.
New sources of CP-violating phases are therefore suppressed.
\end{itemize}
Both types of models will be realized below using a horizontal symmetry
that also generates the SM fermion masses.
As a result, in each case, only a single entry of $y_u$ is $\mathcal{O}(1)$.
This leads to a further suppression of the imaginary parts of
$(\delta_{ij})_{MM}$, since the $\mathcal{O}(1)$ entry only enters
in its absolute value squared.

As mentioned above, the soft mass matrices also contain certain combinations  
of  $y_u$ and the SM Yukawas.
It is easy to verify that these do not introduce any new qualitative 
features in the models considered below.
In the MFV-like models of~\secref{MFV}, $y_u\sim Y_u$, so the discussion above still holds.
In particular, the texture of $y_u$ is unaltered when rotating to the up mass basis.
In the non-MFV models of~\secsref{y11}{y22}, 
these mixed terms are subdominant because there is no
large overlap between $y_u$ and $Y_u$. Such overlap is present in the models
of~\secref{y32}, but as we will see,
flavor constraints are met in these models roughly as discussed above. 
We stress that, for the numerical examples
we present in~\secref{MFV} and~\secref{NonMFV}, 
the full expressions of the soft terms are taken into account.


 \section{MFV-like Models: split top squark with and without charm mixing}
 \label{sec:MFV}

As discussed above, models in which $(y_u)_{ij}$ and $(Y_u)_{ij}$ are the same up 
to $\mathcal{O}(1)$ numbers are consistent with flavor experiments.
This possibility is realized  in models in which  the fermion masses 
are explained by a horizontal symmetry, if the messenger $\bar D$ has the same flavor
charge as $H_U$. 
We thus consider a U(1) horizontal symmetry with a spurion $\lambda$ of charge $-1$,
and matter field charges
 \beq
 Q_1(5),\ Q_2(4),\ Q_3(2),\ u_1(1),\ u_2(-1),\ u_3(-2),\ d_1(1),\ d_2(0),\ d_3(0)\,,
 \eeq
with the Higgses and $\bar D$ messenger neutral.
 The Yukawa matrices are then
 \begin{align}
 y_u &\sim Y_u \sim \begin{pmatrix}
 \lambda^6 & \lambda^4 & \lambda^3\\
 \lambda^5 & \lambda^3 & \lambda^2 \\
 \lambda^3 & \lambda & 1
 \end{pmatrix}\,,
 &
 Y_d &\sim \begin{pmatrix}
 \lambda^6 &\lambda^5 & \lambda^5 \\
 \lambda^5 & \lambda^4 & \lambda^4 \\
 \lambda^3 & \lambda^2 & \lambda^2
 \end{pmatrix}\,,
 \end{align}
where, as explained above, $\sim$ means that the matrices are given up
to ${\cal O}(1)$ coefficients.
These lead to the fermion mixing matrices,
 \begin{align}
 V_{uL} &\sim \begin{pmatrix}
 1 & \lambda & \lambda^3 \\
 \lambda & 1 & \lambda^2 \\
 \lambda^3 & \lambda^2 & 1
 \end{pmatrix}\,,
 &
 V_{uR} &\sim \begin{pmatrix}
 1 & \lambda^{2} & \lambda^3 \\
 \lambda^{2} & 1 & \lambda \\
 \lambda^3 & \lambda & 1
 \end{pmatrix}\,,\nn
 \\
 V_{dL} &\sim \begin{pmatrix}
 1 & \lambda & \lambda^3 \\
 \lambda & 1 & \lambda^2 \\
 \lambda^3 & \lambda^2 & 1
 \end{pmatrix}\,,
 &
 V_{dR} &\sim \begin{pmatrix}
 1 & \lambda & \lambda \\
 \lambda & 1 &1\\
 \lambda &1 & 1
 \end{pmatrix}\,,
 \end{align}
 with $\Vdl \sim \Vul \sim \Vckm$,
 \beq
 \Vckm=\Vul \Vdl^\dag \sim \begin{pmatrix}
 1 & \lambda & \lambda^3 \\
 \lambda & 1 & \lambda^2 \\
 \lambda^3 & \lambda^2 & 1
 \end{pmatrix}.
 \eeq
The corrections to the squark soft mass terms have the structure
 \begin{align}
 \delta\tilde{m}^2_q &\sim \begin{pmatrix}
 \lambda^6 & \lambda^5 & \lambda^3\\
 \lambda^5 & \lambda^4 & \lambda^2 \\
 \lambda^3 & \lambda^2 & 1
 \end{pmatrix}\,,
 &
 \delta\tilde{m}^2_{uR} &\sim \begin{pmatrix}
 \lambda^6 & \lambda^4 & \lambda^3\\
 \lambda^4 & \lambda^2 & \lambda \\
 \lambda^3 & \lambda & 1
 \end{pmatrix}\,,
 \end{align}
 while
 \beq
 A^*_u\sim 
 \begin{pmatrix}
 \lambda^6 & \lambda^4 & \lambda^3\\
 \lambda^5 & \lambda^3 & \lambda^2 \\
 \lambda^3 & \lambda & 1
 \end{pmatrix} \,,
 \eeq
 so a large stop A-term is possible.

Although the above expressions are modified by RGE effects,  
they provide first estimates of the 
flavor-violating terms in the models. 
As discussed in the previous section, the down-RR entries are negligible
in these models. The remaining entries are shown
in~\tableref{boundsMMmodelMFV}. With no CP violating phases,
all of these are significantly below the experimental bounds.
In the presence of CP violation, 
the product  $(\delta^q_{ij})_{LL} (\delta^q_{ij})_{RR}$
is of order the bound quoted in~\tableref{boundsMM}.
However, in the following examples, this product is also below the
bound even for ${\mathcal O}(1)$ phases,
mainly because the typical gluino and average squark masses are higher than 1~TeV.

\begin{figure}[tp]
 \includegraphics[width=0.98\textwidth]{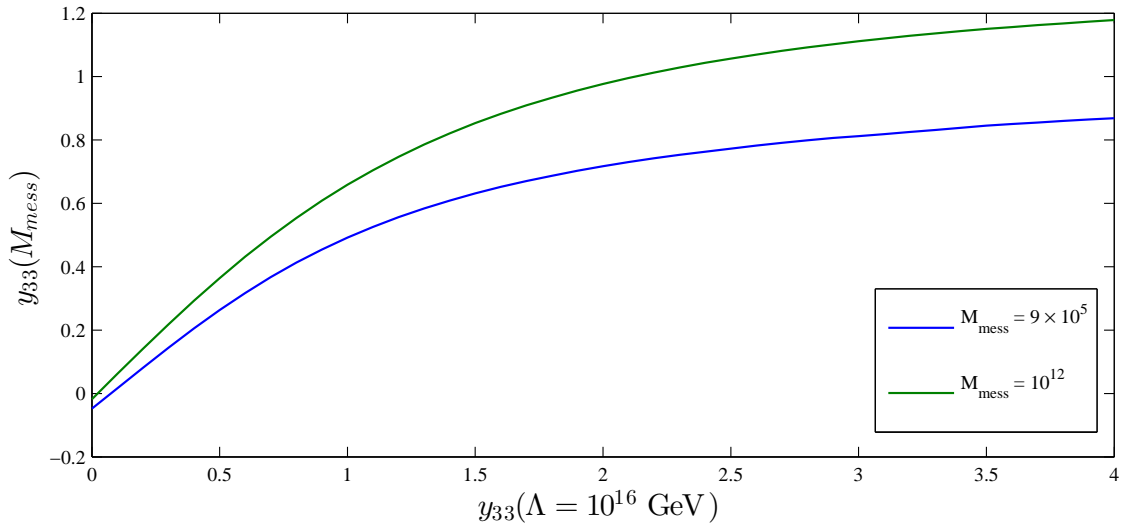}
 \caption{$(y_u)_{33}$ at the messenger scale $M$ as a function of its value at 
the flavor scale $\Lambda$, for $(y_u)_{32}(\Lambda) = 0.2$.}
 \label{fig:vera_yu33_running}
 \end{figure}

\begin{figure}[tp]
 \includegraphics[width=0.98\textwidth]{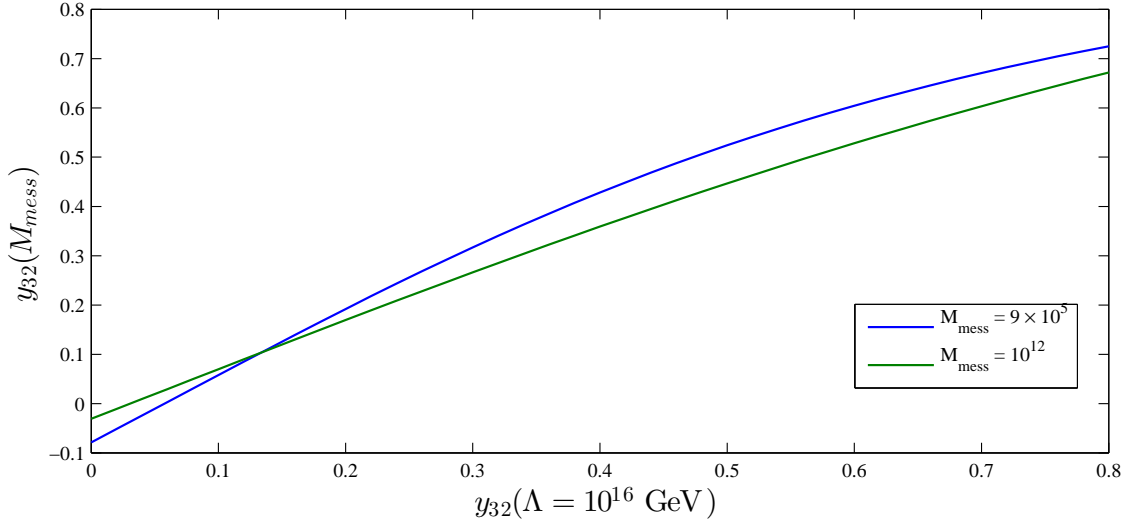}
 \caption{$(y_u)_{32}$ at the messenger scale as a function of $(y_u)_{32}$ at the $\Lambda$ scale, for $(y_u)_{33}(\Lambda) = 1$.}
 \label{fig:vera_yu32_running}
 \end{figure}

%
 \begin{table}[tp]
 \centering
 \renewcommand{\arraystretch}{1.3}
 \begin{tabular}{|cc|ccc|}
 \hline
 $q$ & $ij$ & $| (\delta^q_{ij})_{LL}|$ & $| (\delta^q_{ij})_{RR}|$& 
 $ \sqrt{| (\delta^q_{ij})_{LL}| |(\delta^q_{ij})_{RR}|}$ \\
 \hline
 d & 12 & $\lambda^5$ & $-$ & $-$ \\
u & 12 & $\lambda^5 $ & $\lambda^4 $ & $\lambda^{9/2}$\\
 d & 23 & $\lambda^2 $ & $-$ & $-$\\
 \hline
 \end{tabular}
 \caption{Parametric estimates of $(\delta^q_{ij})_{MM}$ in MFV-like models.
Omitted entries are negligible.}
 \label{tab:boundsMMmodelMFV}
\end{table}
%
%
In both $\delta\tilde{m}^2_q$ and $\delta\tilde{m}^2_{uR}$, 
the dominant contribution is the 3--3 entry. 
The second important contribution is $(\delta\tilde{m}^2_{uR})_{32}$. 
These contributions come from $(y_u)_{33}$ and $(y_u)_{32}$, so the squark spectra 
are particularly sensitive to the ${\cal O}(1)$ coefficients of these two entries.
If the flavor scale, $\Lambda$, is much higher than the the messenger scale $M$, 
these couplings may be significantly modified by the running.
In~\figref{vera_yu33_running}, we show $(y_u)_{33}(M)$ 
as a function of $(y_u)_{33}(\Lambda)$ 
for a given value of $(y_u)_{32}(\Lambda) = 0.2$ and $\Lambda=10^{16}\GeV$.
The value of $(y_u)_{33}$ is clearly decreased by the running.
In particular, large boundary values of $(y_u)_{33}(\Lambda)$, say
above 2, flow to  $(y_u)_{33}\sim1$ at the messenger scale.
A milder decrease, with no IR fixed point behavior, 
is seen in the running of $(y_u)_{32}$, which we display 
in~\figref{vera_yu32_running}.
%
%
%
%

In the following, we discuss two concrete examples of MFV-like models;
one with a low messenger scale, $M= 9.0 \times 10^5\GeV$, 
and the other with a high messenger scale, $M=10^{12}\GeV$. 
The models are obtained by choosing specific ${\cal O}(1)$ coefficients 
of the entries of $y_u$ at the flavor scale $\Lambda=10^{15}\GeV$.
The various couplings are then evolved to the messenger scale $M$,
where the soft masses are calculated and fed as input to  
\softsusy~\cite{Allanach:2001kg},
which is used to obtain the low-energy spectra.

%
 \begin{figure}[tp]
 \includegraphics[width=1\textwidth]{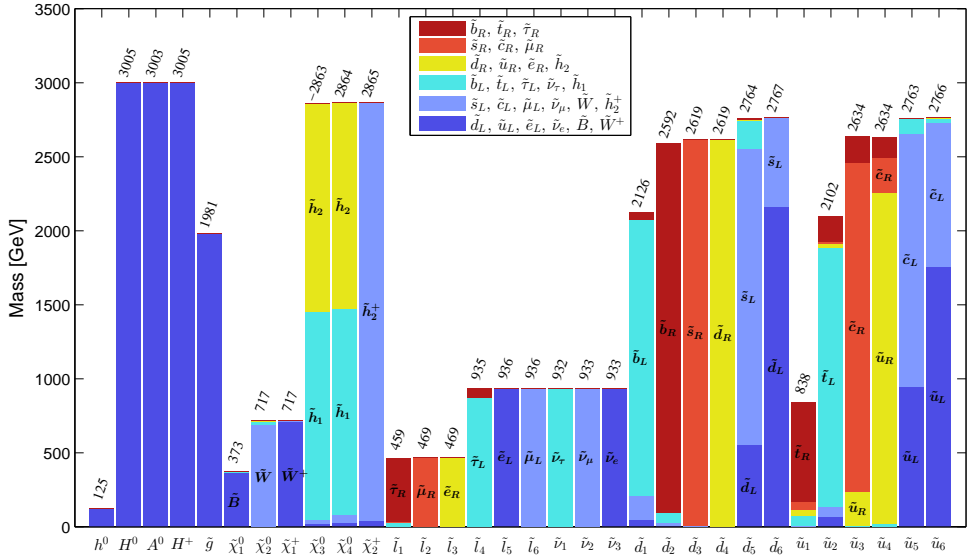}
 \vspace{-2em}
 \caption{Spectrum for BP MFV-t with 
 large $(y_u)_{33}(\Lambda = 10^{15} \GeV) = 2.9$ and scales
 $M = 9 \times 10^{5}$ GeV, $F/M = 2.7 \times 10^5$ GeV.
 Mixings are shown for sfermion and gaugino mass eigenstates.
 In this model, we find $\mu = 2870$ GeV and $(A_u)_{33} = -1920$ GeV.}
 \label{fig:MFV-t}
 \end{figure}
%
\enlargethispage{2\baselineskip}
The first BP ``MFV-t'', where ``t'' represents the stop-like lightest squark, is obtained with the scales $M=9.0\times 10^5\GeV$ 
and $F/M=2.7\times 10^5\GeV$, and is shown in~\figref{MFV-t}.
At the messenger scale, $(y_u)_{33}(M)=0.8$.
The Higgs mass is 125\,GeV, the R-stop is at $\sim 840\GeV$, 
and the L-stop and L-sbottom are near 2.1\,TeV.
The remaining squarks have masses between 2.5 to 2.8 \,TeV, 
and the gluino is at 2~TeV.
The NLSP is the bino, which decays promptly with $c \tau_{\tilde B} > 0.09$~mm
if the $F$ term of $X$ is the dominant $F$-term in the theory.
Thus, the correct Higgs mass is obtained with squarks that are still within 
the reach of the LHC.

From the point of view of mass hierarchies, this spectrum is not unusual,
with the first- and second-generation squarks approximately degenerate;
however, the L-squark mass eigenstates are mixtures of up-charm or down-strange
states, with the size of the mixing of order of the Cabbibo mixing.
Such mixings are completely generic whenever the squarks are not exactly degenerate
because of the fermion Cabbibo mixing.
Note that substantial 1--2 mixing can arise even when the mass splittings are small,
since the leading contribution to the first- and second-generation squark masses
is the flavor-diagonal GMSB contribution~\cite{Feng:2007ke}.
It would be interesting to explore whether this mixing leads to observable effects
at the LHC as charm tagging is improved. 

As noted above,
the only potentially dangerous source of flavor violation
in our examples involves the product of the imaginary parts of the LL 
and RR 1--2 entries in the up sector. 
We list these (for this example and the following examples) 
in Appendix~\ref{sec:app_deltas}.
For the spectrum of \figref{MFV-t}, the bounds of~\tableref{boundsMM}
should be rescaled by $\sim2$. 
The model is thus viable even with ${\mathcal O}(1)$ phases.

%
 \begin{figure}[tp]
 \includegraphics[width=1\textwidth]{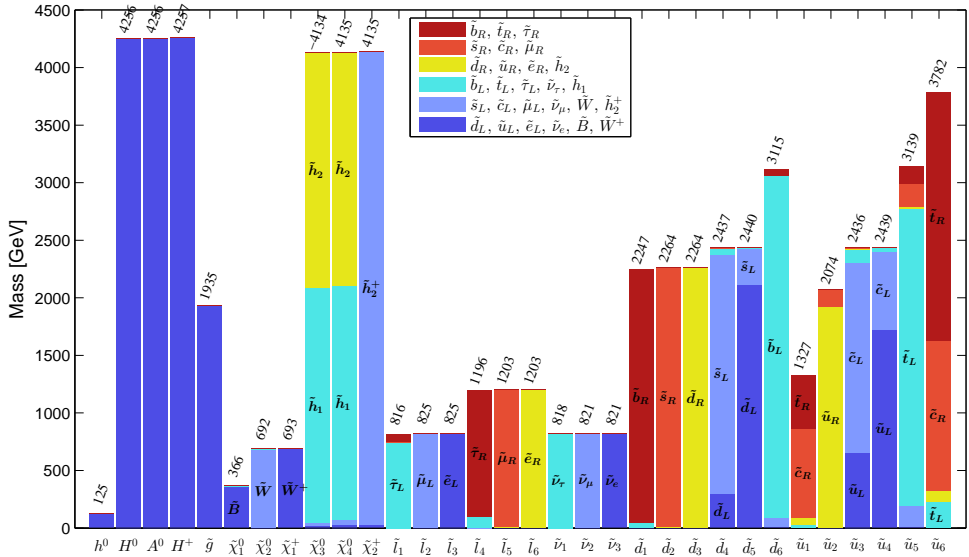}
 \vspace{-2em}
 \caption{Spectrum for BP MFV32 with 
 large $(y_u)_{32}(\Lambda = 10^{15} \GeV) = 0.63$ and scales
 $M = 10^{12}$ GeV, $F/M = 2.71 \times 10^5$ GeV.
 Mixings are shown for sfermion and gaugino mass eigenstates.
 In this model, we find $\mu = 4150$ GeV and $(A_u)_{33} = -2390$ GeV.}
 \label{fig:MFV-ct}
 \end{figure}
%

\enlargethispage{\baselineskip}
In this example, the stop--scharm mixing is quite small.
As discussed above, this mixing is sensitive to the ${\cal O}(1)$
coefficients of the 3--2 entries of $Y_u$ and $y_u$.
The next benchmark point, ``MFV-ct'', features a large stop--scharm mixing 
at a higher messenger scale $M=10^{12}\GeV$, while
keeping $F/M=2.7\times 10^5\GeV$ fixed.
At the messenger scale, $(y_u)_{32}=0.56$, and $(y_u)_{33}=1$.
The resulting spectrum is shown in~\figref{MFV-ct}.
Here, the Higgs mass is 125 GeV. Again, the stop LR mixing, $|X_t|\sim 2.8$~TeV 
gives a large contribution to this mass.
The lightest squark, at 1340 GeV, is an admixture of the R-stop and R-scharm. 
The L-stop and L-sbottom are heavy at $\sim 3.1$~TeV, 
while other squarks are between 2.2--2.5 TeV. 
Due to the high messenger scale, the bino NLSP is long-lived with a lifetime 
of $\tau_{\tilde{B}} \sim 0.3$ s, and does not decay inside the detector.
Note that because of the large stop--scharm mixing, the lightest squark can decay
through either a third- or a second-generation quark, 
thus reducing the sensitivity of stop searches~\cite{Blanke:2013zxo}.

These examples will be probed by the standard LHC supersymmetry searches,
including pair-production of gluinos, squarks, electroweakinos, and stops.
Generally, searches for gluinos using simplified decay chains $\tilde{g} \rightarrow qq\tilde{\chi}_0^1$ 
have bounds of about 2.3 TeV at the high-luminosity LHC \cite{ATL-PHYS-PUB-2014-010, CMS-PAS-SUS-14-012};
in the FGM scenarios, these bounds are weakened by the more complicated cascade decays.
Searches for stop pair-production are particularly relevant for \figref{MFV-t} 
because of the small stop mass \cite{Aad:2015pfx, Khachatryan:2016pup}.
If the bino decays to a $Z$ and a gravitino, and both the $W$ from the top decay and 
the $Z$ decay hadronically, the multi-jet plus missing energy searches are also useful \cite{Aad:2016jxj, Khachatryan:2016kdk}.
In the spectrum of \figref{MFV-ct}, the lightest squark decays into either a second- 
or third-generation quark and the signal strengths of stop searches and
jet plus missing energy squark searches are both reduced.
Finally, searches for direct production of electroweakinos are also relevant for 
bino and wino masses in this range \cite{Aad:2015hea, Khachatryan:2016hns}.


 \section{Non-MFV Models}
 \label{sec:NonMFV}

We now turn to examples with large mass splittings between the
first- and second-generation squarks. 
As explained in \secref{intro}, this requires an alignment
of the down quark and squark mass matrices.
To achieve this, we use a $\mathrm{U}(1) \times \mathrm{U}(1)$ horizontal symmetry. 
We consider three types of examples that realize unusual squark mass 
hierarchies~\cite{Galon:2013jba}.
These examples differ only in the horizontal 
charges of the messenger fields.
The Higgs Yukawa coupling textures are thus the same in the three models, 
but the matter--messenger coupling matrices are different.
Our main focus here is the masses and mixings of the first- and second-generation squarks. 
Thus, we do not insist on a 125\,GeV Higgs.
In models with a lower Higgs mass, as is the case in our first example,
some additional mechanism is required to raise the Higgs mass,
such as the addition of the NMSSM singlet~\cite{Chacko:2001km}.

We choose the quark $\mathrm{U}(1) \times \mathrm{U}(1)$ charges to be,
 \beq
 \begin{split}
 Q&_{1}\left( 3,0\right) ,\,Q_{2}\left( 0,2\right) ,\,Q_{3}\left( 0,0\right),\\
 u&_{1}\left( -3,6\right) ,\,u_{2}\left( 1,0\right) ,\,u_{3}\left( 0,0\right),\\
 d&_{1}\left( -1,4\right) ,\,d_{2}\left( 4,-2\right) ,\,d_{3}\left( 0,2\right).
 \end{split}
 \eeq
We assume that each U(1) is broken by a spurion of charge $-1$ and size $\lambda\sim0.2$.
 These charges give rise to the following structures for the SM Yukawas, up to $\mathcal{O}\left( 1\right) $ coefficients
 \begin{align}
 Y_{u} &\sim \bpm\lambda^{6} & \lambda^{4} & \lambda^{3}\\
 0 & \lambda^{3} & \lambda^{2}\\
 0 & \lambda & 1
 \epm\,,
 &Y_{d} &\sim \bpm\lambda^{6} & 0 & \lambda^{5}\\
 0 & \lambda^{4} & \lambda^{4}\\
 0 & 0 & \lambda^{2}
 \epm\,.
 \end{align}
 This particular structure for $Y_d$ was shown in~\cite{Nir:2002ah} to be the only pattern 
 that can provide alignment in models with horizontal abelian symmetries.\footnote{
This choice of
charges is different from the choice of~\cite{Galon:2013jba}, which gave a zero $V_{td}$.
While the main results of~\cite{Galon:2013jba} still hold, some details of the models
are modified with this choice.}
 The diagonalizing matrices are then
 \begin{align}
 \label{V_matrices}
 \Vul &\sim
 \bpm
 1 & \lambda & \lambda^3 \\
 \lambda & 1 & \lambda^2 \\
 \lambda^3 & \lambda^2 & 1 
 \epm\,,
 &
 \Vur &\sim
 \bpm
 1 & \lambda^4 & \lambda^5 \\
 \lambda^4 & 1 & \lambda \\
 \lambda^9 & \lambda & 1 
 \epm\,,\nn
 \\
 \Vdl &\sim
 \bpm
 1 & \lambda^5 & \lambda^3 \\
 \lambda^5 & 1 & \lambda^2 \\
 \lambda^3 & \lambda^2 & 1 
 \epm\,,
 &
 \Vdr &\sim
 \bpm
 1 & \lambda^7 & \lambda^7 \\
 \lambda^7 & 1 & \lambda^4 \\
 \lambda^7 & \lambda^4 & 1 
 \epm\,,
 \end{align}
which give the correct CKM texture. 
In the following subsections, the single $\mathcal{O}(1)$ entry in $y_u$ will be denoted by $y$.

\subsection{Large $(y_u)_{11}$: light up squark}\label{sec:y11}
For our first example, we take the messenger field to carry 
a horizontal charge of 
 \beq 
\bar{D}\left( 0,-6 \right)\,. 
\eeq
 The messenger-Yukawa matrix is therefore
 \beq
 y_{u} \sim \bpm y & 0 & 0\\
 0 & 0 & 0\\
 0 & 0 & 0
 \epm \,.
 \eeq

The squark mass matrices are schematically given by~\eqref{sqmasses},
with $\Wul \sim \Wur \sim 1_{3\times3}$; the only source of mixing is the fermion
masses. In particular, to a good approximation, the only 1--2 mixing 
is in the L-up sector, as is generically the case with alignment in the down sector,
and the only 2--3 mixing is in the R-up sector; both are ${\cal O}(\lambda)$.
The new coupling mainly affects the up and down squark masses,
and can increase or decrease them depending on the value of $F/M^2$.
The one-loop contributions are negative, while the two-loop
contributions can have either sign.
Since there is no large A-term, the Higgs mass requires extra ingredients,
unless the stops are heavy.

Since $y$ is $\mathcal{O}(1)$, the dominant source of flavor violation is
$\left( \delta^u_{12} \right)_{LL} \sim O(\lambda)$,
which can be consistent with flavor bounds for certain choices of the model parameters.
However, this estimate applies only to the absolute value of 
$\left( \delta^u_{12} \right)_{LL}$.
The contribution to CP-violating processes is further suppressed,
since $y$ only enters the soft masses as $\left| y \right|^2$.
 CP violation therefore originates solely from the SM Yukawas,
\beq
{\rm{Im}} \left( \delta^u_{12} \right)_{LL} \sim {\rm{Im}}
(\Vul\, \yud\, \yud^\dagger\,\Vul^\dagger)_{12}=
\vert y\vert^2  (\Vul)_{11} {\rm{Im}}(\Vul^*)_{21} \sim \lambda^5\,,
\eeq
since $\Vur\sim \Vckm$ in the 1--2 block.
The imaginary part of $\left( \delta^u_{12} \right)_{RR}$
also originates from the SM Yukawas, but in this case, its source is $\Vur$,
\beq
{\rm{Im}} \left( \delta^u_{12} \right)_{RR} \sim {\rm{Im}}
(\Vur\, \yud^\dagger\, \yud\,\Vur^\dagger)_{12}=
\vert y\vert^2  (\Vur)_{11} {\rm{Im}}(\Vur^*)_{21}\,.
\eeq
Thus,  $\vert  \left( \delta^u_{12} \right)_{RR}  \vert \sim \lambda^4$,
and is very small.
Still,
$\sqrt{{\rm Im}((\delta^u_{12})_{LL} (\delta^u_{12})_{RR})}$ may be above the
bound of \tableref{boundsMM} by a factor of a few, and viable models may 
require some suppression of  the phases in the 1--2 block of $\Vur$.
In the examples below, the required suppression is $\sim0.4$.

\begin{figure}[tp]
 \includegraphics[width=1\textwidth]{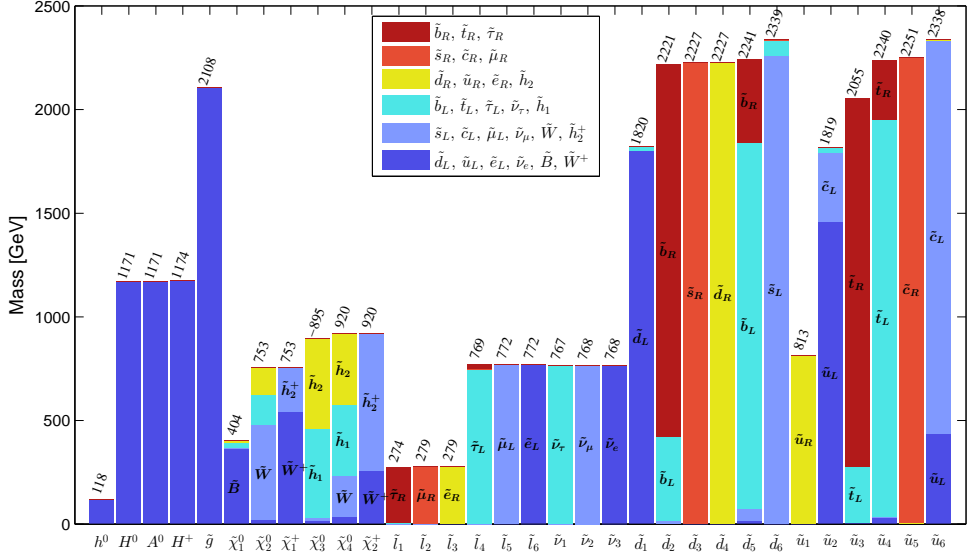}
 \vspace{-2em}
 \caption{Spectrum for BP U11 with 
 $N = 2$, large $(y_u)_{11}(\Lambda = 10^{16}\GeV) = 0.4$,
 and scales $M = 6.0 \times 10^{5}$ GeV, $F/M = 1.49 \times 10^5$ GeV.
 Mixings are shown for sfermion and gaugino mass eigenstates.
 In this model, we find $\mu = 885$ GeV and $(A_u)_{33} = -532$ GeV.}
  \label{fig:U11}
 \end{figure}

\Figref{U11} shows the spectrum of a BP, ``U11'', with $N_5=2$. 
Here the gluino is at 2.1~TeV. The R-up squark is at 810~GeV,
while all other squarks have masses between 1.8--2.3~TeV. 
%
%
Thus, there is a single light squark in the spectrum.
The large squark mass splitting gives a negative (positive) contribution to
the R(L)-slepton masses through the running~\cite{Chacko:2001km,Abdullah:2012tq,Evans:2012hg}
resulting in an R-stau NLSP with a
decay length of $c \tau_{\tilde\tau} \geq 5.0$ mm,
where the lower value is attained if the dominant
source of supersymmetry breaking is the $F$-term of $X$. 
For a larger value of this parameter, the stau can become long-lived 
and detectable in searches for long-lived charged particles \cite{ATLAS:2014fka, Chatrchyan:2012jwg}.

 \subsection{Large $(y_u)_{22}$: light charm squark}\label{sec:y22}

\enlargethispage{3\baselineskip}
This example is qualitatively similar, with the large effects occurring in 
the second-generation squarks.
We take the $\bar D$ messenger charges to be,
 \beq 
\bar{D}\left( -1,-2\right)\,, 
\eeq
so that
 \beq y_{u} \sim \bpm0 & 0 & 0\\
 0 & y & 0\\
 0 & 0 & 0
 \epm \,.
\eeq
The main effects are therefore on the charm and strange squark masses,
and these can have either sign.

We note that the horizontal charges also allow a small $O(\lambda^3)$ mixing between 
$H_U$ and $\bar D$ from the \kahler and superpotential, changing the Yukawa structure to (see Appendix~\ref{sec:fgmapp})
 \beq
 y_u \rightarrow \bpm \sim 0 & \lambda^7 & \lambda^6 \\
 0 & y & \lambda^5\\
 0 & \lambda^4 & \lambda^3
 \epm \,.
 \eeq
These corrections are small, and proportional to the SM Yukawas.
Therefore the squark--quark mixings are again given by the SM flavor mixing 
matrices, with $\Wul=\Wur\sim1_{3\times3}$, as in the previous example. 

\begin{figure}[tp]
 \includegraphics[width=1\textwidth]{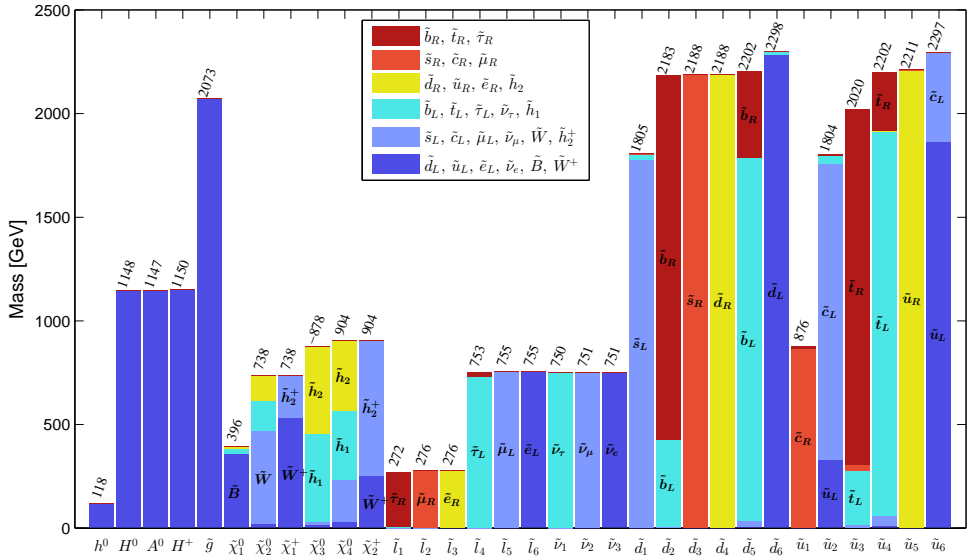}
 \vspace{-2em}
 \caption{Spectrum for BP U22L with $N_5 = 2$,
 $M = 5.7 \times 10^{5}\GeV$, $F/M = 1.46 \times 10^5 \GeV$, and
 $(y_u)_{22}(\Lambda = 10^{16}\GeV) = 0.7$. 
 Mixings are shown for sfermion and gaugino mass eigenstates
 In this model, we find $\mu = 869\GeV$ and $A_u^{33} = -521\GeV$.}
 \label{fig:U22L}
 \end{figure}

We show two examples in which the new contributions to the charm squark
are negative, resulting in a single charm squark that is much lighter than
the remaining squarks. The flavor-violating $\delta$'s in these examples 
are similar to those of the previous example.
In~\figref{U22L} we display the spectrum for BP ``U22L'' obtained for a messenger scale 
$M = 5.7 \times 10^5\GeV$ and $N_5=2$. 
%
%
%
 \begin{figure}[tp]
 \includegraphics[width=1\textwidth]{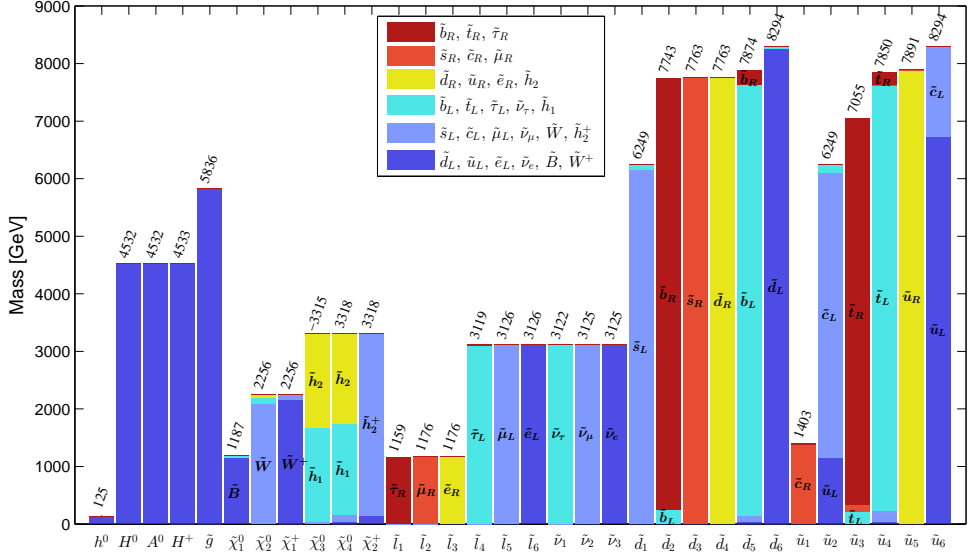}
 \vspace{-2em} 
 \caption{Spectrum for BP U22 with
 large $(y_u)_{22}(\Lambda = 10^{16}\GeV) = 0.73$ and scales
 $M = 10^{7}$ GeV, $F/M = 8.6 \times 10^5$ GeV.
 Mixings are shown for sfermion and gaugino mass eigenstates.  
 In this model, we find $\mu = 3280$ GeV and $(A_u)_{33} = -1540$ GeV.}
 \label{fig:U22}
 \end{figure}
%
The R-scharm is at 870~GeV, the remaining squarks have masses above 1.8~TeV, 
and the gluino mass is around 2~TeV.
Here and in the next BP U22,
the large squark mass splitting feeds into the RGE's of the sleptons 
and creates a large splitting between the R- and L-sleptons,
so that the R-stau is the NLSP, while the L-sleptons are all heavier than the winos.
The stau NLSP decays to a tau and gravitino, with a lower bound on the decay length of 
$c \tau_{\tilde \tau} \geq 0.05$~mm;
for much longer lifetimes, 
both examples would be excluded by long-lived stau searches.
The R-sleptons have approximately degenerate masses of $\sim 270$ GeV, 
which currently lie above searches using direct production \cite{Aad:2014vma, Khachatryan:2014qwa}.
SUSY searches with multilepton final states can also be useful to look for
cascade decays of the R-up squark \cite{Aad:2015wqa}.
In contrast to minimal GMSB spectra, the quarks are almost exclusively of charm flavor, 
so charm tagging will play an important role in these cases \cite{Aad:2015gna}.

The second BP ``U22'' demonstrates that a large Higgs mass can be obtained
in these models from very heavy stops, while a single squark may still 
be within LHC reach. The stops in this example are at 7~TeV.
Since only a single squark flavor is significantly affected by the messenger coupling,
most of the remaining squarks are around 7~TeV too, with the exception of the R-scharm at 1.4~TeV. 
The spectrum is shown in~\figref{U22}, for $M = 10^7\GeV$, 
$y(\Lambda = 10^{16}\GeV) = 0.73$, and $F/M = 8.6 \times 10^5\GeV$.

This spectrum has a long-lived NLSP stau with $c\tau_{\tilde\tau} \ge 0.4$ mm,
which is mainly produced at the LHC in the cascade decay of $\tilde u_1$. 
Therefore, we expect two long-lived charged particles accompanying two quarks.

\subsection{Large $(y_u)_{32}$: light top and/or charm squark}\label{sec:y32}
 
\enlargethispage{3\baselineskip}
As a final example, we consider models with a large 3--2 entry in the messenger coupling matrix.
This leads to a large stop A-term, which allows for a 125 GeV Higgs with 
superpartners accessible at the LHC. 
At the same time, there are large modification
of the L-stop, L-sbottom and R-scharm masses, and in some examples, 
a large stop--scharm mixing.  

To obtain a large $(y_u)_{32}$ we take the messenger horizontal charges to be
\beq 
\bar{D}\left( -1,0\right)\,, 
\eeq
which yields the messenger Yukawa matrix
 \beq
 y_{u} \sim \bpm 0 & \lambda^{3} & \lambda^{2}\\
 0 & \lambda^{2} & 0\\
 0 & y & 0
 \epm \,.
 \eeq
As above, $y$ is an ${\cal O}(1)$ coefficient which we display explicitly 
since it is the dominant entry in the $y_u$ matrix.

As discussed in~\appref{fgmapp}, these charges also allow a $\lambda X D H_U$ 
term in the superpotential, and a $\lambda \bar D^\dag H_U$ term in the \kahler 
potential. Therefore, the Yukawa matrices may in principle be modified as  
$Y_u \rightarrow Y_u + \lambda y_u$ and $y_u \rightarrow y_u + \lambda Y_u$.
This only affects the texture of $y_u$, with
\begin{align}
 \label{model32_highscale_yukawas}
y_{u} \to\bpm \lambda^7 & \lambda^{3} & \lambda^{2}\\
 0 & \lambda^{2} & \lambda^3\\
 0 & y & \lambda
 \epm\,.
 \end{align} 

Since the messenger coupling has an $\mathcal{O}(1)$ 3--2 entry,
mixed terms involving both $y_u$ and the SM Yukawas are important in these models.
Examining the structure of the soft terms in~\eqref{app_sqmasses},
one gets,
\begin{align}
 \begin{split}
 \delta \tilde m^2_q \sim & -\frac{1}{(4\pi)^2} \frac{1}{6}
 \bpm
 \lambda^4 & \lambda^5 & y\lambda^3 \\
 \lambda^5 & \lambda^4 & y\lambda^2 \\
 y\lambda^3 & y\lambda^2 & |y|^2 
 \epm \frac{F^4}{M^6}
 \\&+ \frac{1}{(4\pi)^4}
 \bpm
 (3|y|^2 + 2 - G) \lambda^4 & (6|y|^2 + 3y +2 - G)\lambda^5 & (6y^3 +5y -Gy) \lambda^3 \\
 (6|y|^2 + 3y +2 - G)\lambda^5 & (6|y|^2 - G) \lambda^4 & (6y^3 - Gy) \lambda^2 \\
 (6y^3 +5y -Gy) \lambda^3 & (6y^3 - Gy) \lambda^2 & (6|y|^4 - G|y|^2) 
 \epm \left|\frac{F}{M}\right|^2 \,,
 \end{split}
\\
 \begin{split}
 \delta \tilde m^2 _{u_R} \sim & -\frac{1}{(4\pi)^2} \frac{1}{3}
 \bpm
   0 &   0 &   0 \\
   0 & |y|^2 & -y \lambda^3 \\
   0 & -y \lambda^3 & \lambda^4 
 \epm \frac{F^4}{M^6}
 \\&+ \frac{1}{(4\pi)^4}
 \bpm
   0 &   0 &   0 \\
   0 &(2-2G)|y|^2 + 12 |y|^4 & 4 |y|^2 \lambda \\
   0 & 4|y|^2 \lambda & -2|y|^2
 \epm \left|\frac{F}{M}\right|^2 \,,
 \end{split}
 \\
 \delta \tilde m^2_{d_R} \sim & -\frac{1}{(4\pi)^2} 2
 \bpm
   0 &   0 &   0 \\
   0 &   0 &   0 \\
   0 &   0 & |y|^2 \lambda^4
 \epm \left|\frac{F}{M}\right|^2 \,, \\
 A_u \sim & -\frac{1}{(4\pi)^2}
 \bpm
   0 & (y+2|y|^2) \lambda^4 & y \lambda^3 \\
   0 & (y+2|y|^2) \lambda^3 & y \lambda^2 \\
   0 & 3|y|^2 \lambda & |y|^2
 \epm \frac{F}{M} \,,
 \\
 A_d \sim & -\frac{1}{(4\pi)^2}
 \bpm
   0 &   0 & y \lambda^5 \\
   0 &   0 & y \lambda^4 \\
   0 &   0 & |y|^2 \lambda^2
 \epm \frac{F}{M} \,,
 \end{align}
where we neglected ${\mathcal O}(\lambda^6)$ terms.
\enlargethispage{2\baselineskip}
 Here $G = \frac{16}{3}g_3^2 + 3g_2^2 +\frac{13}{15}g_1^2$ is between 
5 and 10, depending on the messenger scale. Recall that, in addition to $G$ 
and the explicit numerical factors that appear above, 
there are other unknown numerical factors coming from the 
Yukawa couplings themselves.
The quark-squark gluino mixings are then given by,
\begin{align}
 K_u^L &\sim 
 \bpm
 1 & \lambda & \lambda^3 \\
 \lambda & 1 & \lambda^2 \\
 \lambda^3 & \lambda^2 & 1 
 \epm \,,
 &
 K_u^R &\sim 
 \bpm
 1 & \lambda^4 & \lambda^5 \\
 \lambda^4 & 1 & \lambda \\
   0 & \lambda &1 
 \epm \,,
 \nn\\
 K_d^L &\sim 
 \bpm
 1 & \lambda & \lambda^3 \\
 \lambda & 1 & \lambda^2 \\
 \lambda^3 & \lambda^2 & 1 
 \epm \,,
 &
 K_d^R &\sim 
 \bpm
 1 & \lambda^3 & \lambda^7 \\
 \lambda^3 & 1 & \lambda^4 \\
 \lambda^7 & \lambda^4 & 1 
 \epm \,.
 \end{align}
These models then have the following general features:
 \begin{enumerate}
 \item A large negative contribution to the R-stop mass 
$\left( \tilde m^2_{u_R} \right)_{33}$. 
 \item Large contributions to the R-scharm mass, $\left( \tilde m^2_{u_R} \right)_{22}$.
Here the one-loop contribution is negative, 
and the two-loop contribution can have either sign.
 \item Large contributions to the L-stop and L-sbottom masses, 
$\left( \tilde m^2_q \right)_{33}$. 
Here the one-loop contribution is negative, while
the two-loop contribution can have either sign.
 \item An $\mathcal{O}(\lambda)$ R stop--scharm mixing, $\left( \tilde m^2_{u_R} \right)_{32}$.
 \item A large stop A-term $\left( A_u \right)_{33}$.
 \end{enumerate}
The parametric estimates for the relevant flavor-violating quantities in these models
are collected in~\tableref{boundsMMalign32}.
These are clearly compatible with current experimental bounds.
In particular, $(\delta^q_{12})_{LL}$ is small mainly because of the ${\cal O}(\lambda^4)$ mass
splitting in the L-squarks (accompanied by an ${\cal O}(\lambda)$ mixing). 
In contrast, there is an ${\cal O}(1)$ mass splitting
between the R-charm and R-up squarks, but the mixing between them is ${\cal O}(\lambda^4)$.
%
\begin{table}[tb]
 \centering
 \renewcommand{\arraystretch}{1.3}
 \begin{tabular}{|cc|ccc|}
 \hline
 $q$ & $ij$ & $| (\delta^q_{ij})_{LL}|$ & $| (\delta^q_{ij})_{RR}|$& 
 $ \sqrt{| (\delta^q_{ij})_{LL}| |(\delta^q_{ij})_{RR}|}$ \\ \hline
 $d$ & 12 & $\lambda^5$ & $-$ & $-$\\
 $u$ & 12 & $\lambda^5$ & $\lambda^4$ & $\lambda^{4.5}$ \\
 $d$ & 23 & $\lambda^2$ & $\lambda^8$ & $\lambda^5$\\
 \hline
 \end{tabular}
 \caption{Parametric estimates of $(\delta^q_{ij})_{MM}$ in non-MFV models
 with large $(y_u)_{32}$.
Omitted entries are negligible.}
 \label{tab:boundsMMalign32}
 \end{table}
%

\begin{figure}[tp]
 \centering
 \includegraphics[width=1\textwidth]{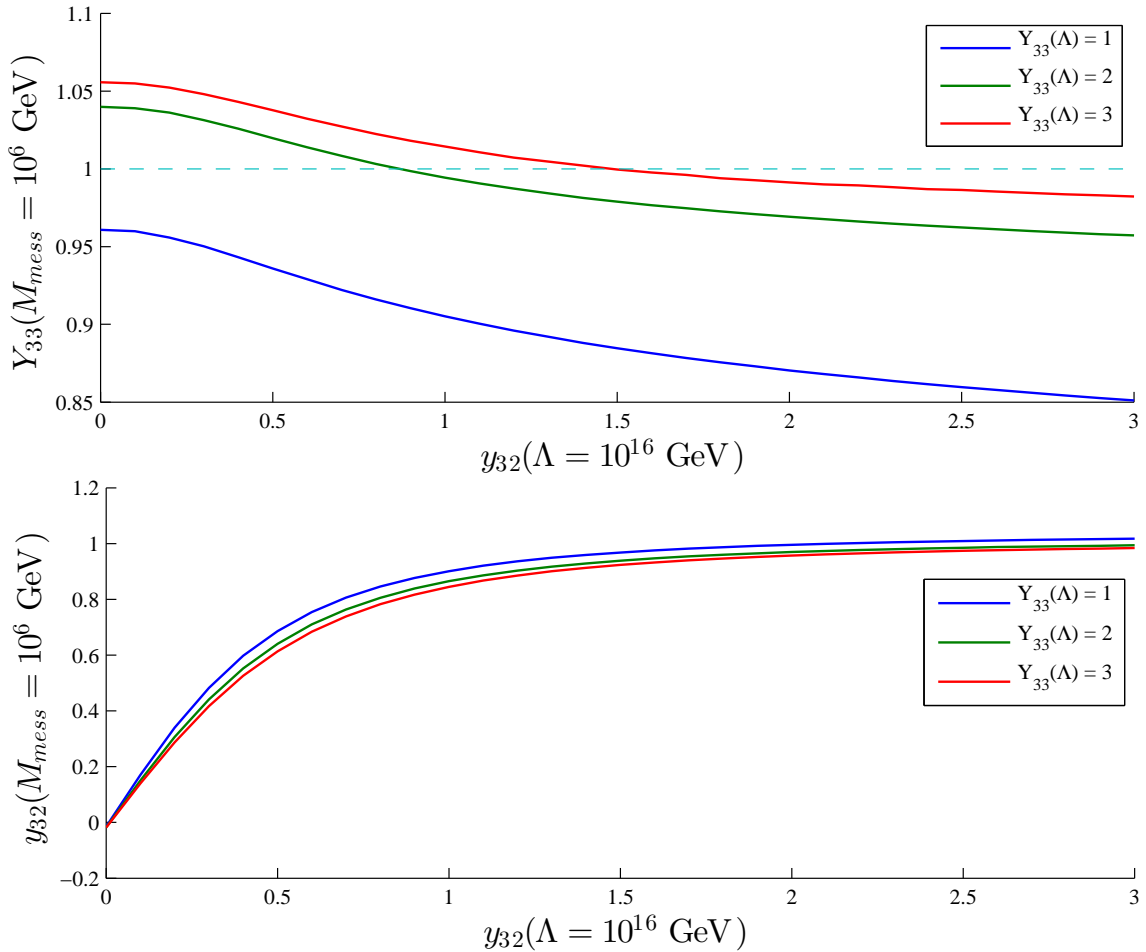}
 \caption{$(y_u)_{32}$ and $(Y_u)_{33}$ at the messenger scale $M = 10^{6}\GeV$ as functions of $(y_u)_{32}(\Lambda) = y$ 
 for various $\mathcal{O}(1)$ values of $(Y_u)_{33}(\Lambda)$.}
 \label{fig:model32_high_low}
 \end{figure}

As in the MFV-like models, since the messenger couplings to the third
generation are large here, the running between the flavor scale $\Lambda$ and the messenger
scale $M$ can modify both the messenger couplings and the SM Yukawa couplings.
In~\figref{model32_high_low} we show $(Y_u)_{33}(M)$ and $(y_u)_{32}(M)$ 
as a function of $(y_u)_{32}(\Lambda) = y$, for various values of $(Y_u)_{33}(\Lambda)$.
As it decreases in the running, $(Y_u)_{33}$ requires a large boundary value 
at $\Lambda$.
Furthermore, for a wide range of boundary values at $\Lambda$, 
$(y_u)_{32}$  flows to values near 1 at the messenger scale.
Thus the size of the new messenger coupling is limited by the running.
%
%

\begin{figure}[tp]
 \includegraphics[width=1\textwidth]{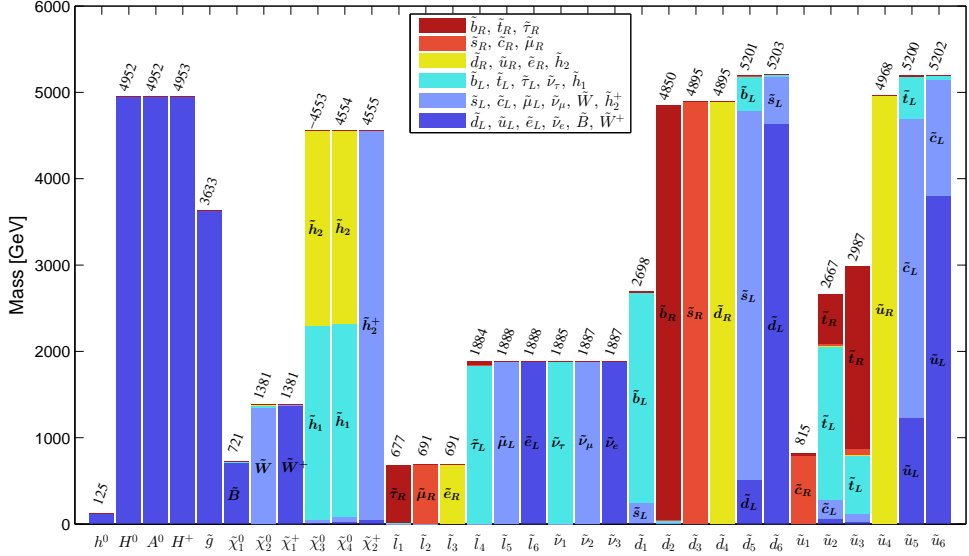}
 \vspace{-2em} 
 \caption{Spectrum for BP U32-c with
 large $(y_u)_{32}(\Lambda = 10^{16}\GeV) = 0.89$
 and scales $M = 3.0 \times 10^{6}$ GeV, $F/M = 5.25 \times 10^5$ GeV.
 Mixings are shown for sfermion and gaugino mass eigenstates.
 In this model, we find $\mu = 4580$ GeV and $(A_u)_{33} = -2880$ GeV.}
 \label{fig:U32-c}
 \end{figure}

\begin{figure}[tp]
 \includegraphics[width=1\textwidth]{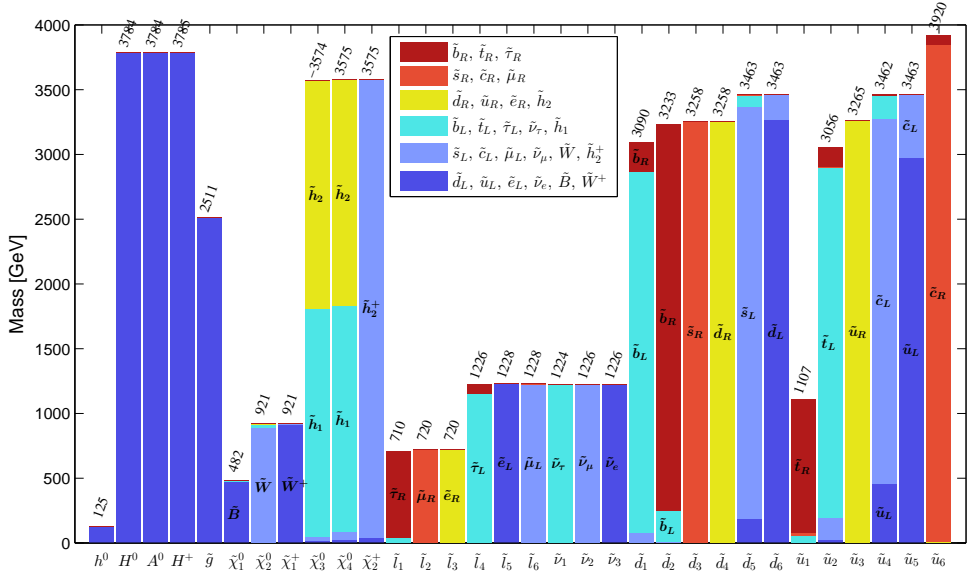}
 \vspace{-2em}
 \caption{Spectrum for BP U32-t with
 $M = 10^{7}\GeV$, $F/M = 3.52 \times 10^5 \GeV$, and
 $y_u^{32}(\Lambda = 10^{16}\GeV) = 1.57$. 
 Mixings are shown for sfermion and gaugino mass eigenstates.
 In this model, we find $\mu = 3580\GeV$ and $A_u^{33} = -2360\GeV$.}
 \label{fig:U32-t}
 \end{figure}

\begin{figure}[tp]
 \includegraphics[width=1\textwidth]{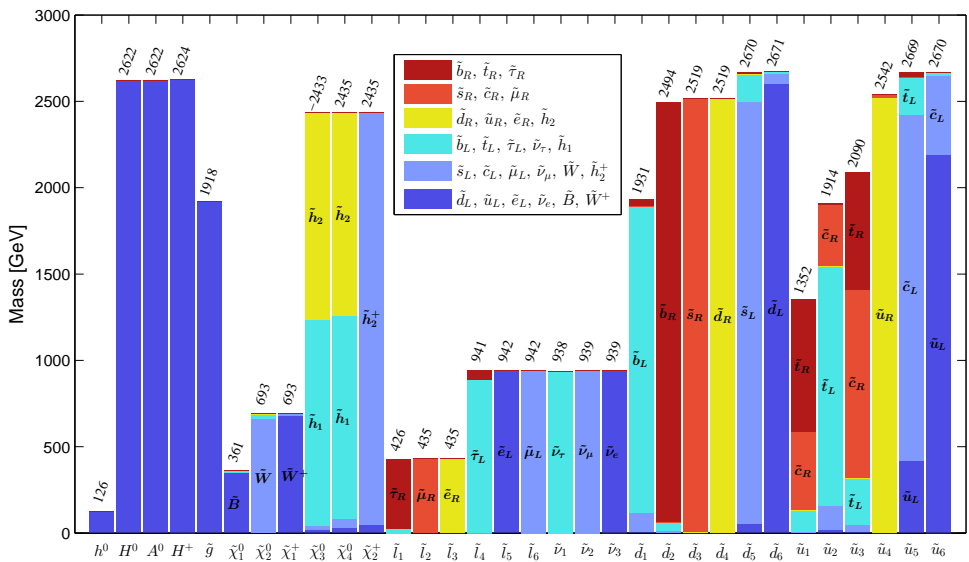}
 \vspace{-2em}
 \caption{Spectrum for BP U32-ct with
 large $(y_u)_{32}(\Lambda = 10^{16}\GeV) = 1.21$, $(y_u)_{33}(\Lambda = 10^{16}\GeV) = 0.6$
 and scales $M = 3.0 \times 10^{6}$ GeV, $F/M = 2.66 \times 10^5$ GeV.
 Mixings are shown for sfermion and gaugino mass eigenstates.
 In this model, we find $\mu = 2450$ GeV and $(A_u)_{33} = -1460$ GeV.}
 \label{fig:U32-ct}
 \end{figure}

\pagebreak
Three spectra of benchmark points using this model are shown in 
Figs.~\ref{fig:U32-c}--\ref{fig:U32-ct}.
All of these have a Higgs mass of 125--126 GeV, partly driven by the large stop mixing,
which in turn comes from the large stop A-term coupled with, in some cases,
a negative contribution to the stop soft masses from $(y_u)_{32}$.
We have taken the flavor scale $\Lambda = 10^{16}$~GeV in these examples. 

%
%
The first BP ``U32-c'' features a very heavy spectrum, with a single squark below the TeV scale.
Specifically, the R-scharm is at 815~GeV,
the stops and L-sbottom are between 2.7--3~TeV,
and all other colored superpartners are above 4~TeV.
This spectrum, shown in~\figref{U32-c} is obtained for 
$M = 3.0 \times 10^6\GeV$, $(y_u)_{32}(\Lambda) = 0.89$, 
$F/M = 5.25 \times 10^5\GeV$, 
where the one- and two-loop contributions are comparable. 
The NLSP is the stau; it promptly decays to a tau and gravitino
with the lower bound $c \tau_{\tilde \tau} > 5$ mm.

\enlargethispage{\baselineskip}
In the next spectrum for BP ``U32-t'', shown in \figref{U32-t}, 
the two-loop contribution is dominant for the parameter choices
$M = 10^7\GeV$, $F/M = 3.52 \times 10^5\GeV$, 
and $(y_u)_{32}(\Lambda) = 1.57$.
The gluino is at 2.5~TeV and all squarks, with the exception of the R-stop, 
are between 3--4~TeV. Due to the negative contribution from $(y_u)_{32}$,
the mass of the R-stop is much lower at 1.1~TeV.
The NLSP in this model is the bino, with a lower bound on the decay length of 
$c \tau_{\tilde B} > 4$ mm.

%
%

\enlargethispage{\baselineskip}
Finally, the last BP ``U32-ct'', shown in~\figref{U32-ct},
has a lighter spectrum, but features a mixed scharm--stop as the lightest
squark around 1.3~TeV, making stop searches relying on decays to 
third-generation quarks even more challenging. 
The lower bound on the decay length of the bino NLSP is
$c \tau_{\tilde B} > 0.04$ mm.

The lightest squark and NLSP of these models will largely determine their collider signatures.
In terms of relevant searches, each of these examples has overlap with previous examples:
\figref{U32-c} with \figref{U22} (light R-scharm and R-stau NLSP),
\figref{U32-t} with \figref{MFV-t} (light R-stop and bino NLSP), and
\figref{U32-ct} with \figref{MFV-ct} (light mixed R stop--scharm and bino NLSP).
The NLSP lifetimes for Figs. 9--10 are longer due to the larger values of $F$.
Comparing the flavour bounds in the 1--2 up sector listed in \tableref{deltau12},
we see that the constraints from observables in D mesons can be used to distinguish these pairs of spectra.

%
%

Note that the models discussed in this subsection have a non-trivial A-term structure,
with a potentially large $(A_u)_{32}$. Furthermore, some of the examples
have a large stop--scharm mixing. Both of these features may affect the 
calculation of the Higgs mass by a few 
GeV~\cite{Cao:2006xb,Kowalska:2014opa,Brignole:2015kva,Goodsell:2015yca}.
These effects are not taken into account in~\softsusy.


 \section{Conclusions}
\label{sec:concl}
We presented examples of flavored gauge mediation 
with interesting and unusual squark spectra that are nonetheless consistent
with low-energy constraints.

These examples are derived from fully calculable models.
Specifically, the soft terms are generated by messenger-field loops,
and the size of the different superpotential couplings---both the SM Yukawas
and the messenger Yukawas---are determined by a flavor symmetry.
In many cases, the structure of the soft terms leads to stronger
suppression of low-energy flavor-violating processes, compared to
naive estimates which are based on the flavor symmetry alone.

The smallness of observed flavor-violating processes has
long been viewed as a major constraint on superpartner flavor.
The above discussion suggests that this viewpoint is perhaps too restrictive,
and is the consequence of considering general ansatze for supersymmetry 
and for flavor, rather than concrete models.


\section*{Acknowledgments}
We thank Iftah Galon, Mark Goodsell,  Sven Heinemeyer, Yossi Nir, 
Gilad Perez, and Ofri Telem for discussions.
Research supported by the Israel Science Foundation (Grant No.~ 720/15), by the United-States-Israel Binational Science Foundation (BSF) (Grant No.~2014397),
and by the ICORE Program of the Israel Planning and Budgeting Committee (Grant No. 1937/12).


 \appendix
\section{FGM superpotential and symmetries}
\label{sec:fgmapp}
We summarize some details of the construction of the models from~\cite{Shadmi:2011hs,Galon:2013jba}.
The form of the superpotential in ~\eqref{superpot} can be enforced by various choices
of symmetries. 
Clearly, the messengers are taken to have the same R-parity as the Higgs
fields.
One possibility (which is not necessarily the most economical one) 
is to introduce a U(1)$_X$ under which $X$ has charge 1,
and $T_I$,  $D_{I\neq2}$, $\bar D_2$ have charge $-1$,
as well as a messenger parity for the messenger fields with $I>2$.\footnote{
Motivated by the fact that full dynamical supersymmetry-breaking models that
generate the required vacuum expectation values for $X$ often require an
$X^3$ term~\cite{Dine:1995ag}, earlier FGM papers starting 
with~\cite{Shadmi:2011hs} considered a $Z_3$ instead of a U(1).}
This still allows for the superpotential terms $X D_1 H_U$  and $X \bar D_2 H_D$.
We return to these below.

To allow only up-type messenger couplings, we take instead  $T_I$ and $D_I$ to have
U(1)$_X$ charge $-1$. The only term we can add to the superpotential in \eqref{superpot}
is $\gamma X D_1 H_U$, where $\gamma$ is some coefficient.
In the MFV-like models, where the flavor symmetry does not distinguish between $\bar D_1$ and  
$H_U$, $\gamma\sim1$, and we can redefine the messenger field as the combination of $H_U$
and $\bar D_1$ that couples to $X$, thus eliminating the term $\gamma X D_1 H_U$.

In the non-MFV models of this paper, $\gamma$ is given by some
power of $\lambda$. In addition, the K\"ahler potential
contains mixing terms such as $\delta\bar D^\dag H_U$ with $\delta\sim\gamma$.
Redefining and rotating the fields to get a canonical K\"ahler
potential and to eliminate the superpotential term $X D_1 H_U$,
one obtains  the superpotential in \eqref{superpot} with
$Y_U \rightarrow Y_U + \gamma y_U$, and $y_U \rightarrow y_U + 
\epsilon Y_U$, where $\epsilon\sim\gamma,\delta$.
Alternatively, one can use the flavor symmetry to eliminate the superpotential mixing term
altogether~\cite{Shadmi:2011hs}.

We also assume that the $\mu$ term is forbidden by an additional PQ symmetry.
We will not discuss the origin of the $\mu$ term, 
although it is possible to embed this PQ symmetry in
the flavor symmetry~\cite{Shadmi:2011hs}, and to generate a small ``supersymmetric'' 
$\mu$ term.\footnote{We note that many of our examples require only a small $B$ term
which can be obtained radiatively from the $\mu$ term.}

 \section{Soft Terms}
 \label{sec:soft}

We first specify our conventions for the Yukawas and soft terms.
We define the Lagrangian and the superpotential as
\begin{align}
 \mathcal L &= \int\dd^4\theta\,K + \left(\int\dd^2\theta\,W + \text{h.c.}\right),\\
 W &= X(T_I \bar T_I + D_I \bar D_I)
      + y_{Uij}\bar D_1 q_i u^c_j
      + Y_{Uij} H_U q_i u^c_j
      + Y_{Dij} H_D q_i d^c_j
      + Y_{Lij} H_D l_i e^c_j,
\end{align}
which yields the SM Yukawa interactions
\begin{equation}
 \mathcal L \supset  - Y_{Uij} H_U (\psi_q)_i (\psi_{u^c})_j \,,
\end{equation}
etc., and the soft terms
\begin{equation}
 -\mathcal L \supset
      \tilde m^2_{qij} \tilde q^*_i \tilde q_j
      + \tilde m^2_{lij} \tilde l^*_i \tilde l_j
      + \tilde m^2_{u_Rij} \tilde u^*_i \tilde u_j
      + \tilde m^2_{d_Rij} \tilde d^*_i \tilde d_j + \cdots
      + A_{uij}H_U \tilde q_i \tilde u_j^* + \cdots,
\end{equation}
where $\tilde q$ and $\psi_q$ ($\tilde u^*$ and $\psi_{u^c}$) are the scalar and fermion components 
of the superfield $q$ ($u^c$). In this paper, following \cite{Abdullah:2012tq}, 
we use the common notation for the SM Yukawas
\begin{equation}
 Y_u = (Y_U)^*,\qquad
 Y_d = (Y_D)^*,\qquad
 y_u = (y_U)^*,
\end{equation}
which appears in the Lagrangian as (e.g., Sec. 11 of \cite{Beringer:1900zz}),
\begin{equation}
 \mathcal L \supset - Y_{uij}H_U(\overline{\psi_q})_i(\psi_u)_j.
\end{equation}

 The leading order GMSB contributions to the soft masses 
 are \cite{Dine:1993yw,Dine:1994vc,Dine:1995ag}
 \begin{align}
 \tilde M_i &= \frac{g_i^2}{(4 \pi)^2} N_5 \frac{F}{M} g(x) \,,
 \\
 \tilde m^2_{H_U} &= \frac{1}{(4\pi)^4} 2N_5\left( \frac{3}{4} g_2^4 + \frac{3}{20}g_1^4 \right) \left|\frac{F}{M}\right|^2\,,
 \\
 \tilde m^2_{H_D} &= \frac{1}{(4\pi)^4} 2N_5\left( \frac{3}{4} g_2^4 + \frac{3}{20}g_1^4 \right) \left|\frac{F}{M}\right|^2\,,
 \\
 \tilde m^2_q &= \frac{1}{(4\pi)^4} 2N_5\left( \frac{4}{3}g_3^4 + \frac{3}{4}g_2^4 + \frac{1}{60}g_1^4\right) \left|\frac{F}{M}\right|^2 1_{3\times 3} \,,
 \\
 \tilde m^2_{u_R} &= \frac{1}{(4\pi)^4} 2N_5\left( \frac{4}{3}g_3^4 + g_1^4\frac{4}{15}\right) \left|\frac{F}{M}\right|^2 1_{3\times 3} \,,
 \\
 \tilde m^2_{d_R} &= \frac{1}{(4\pi)^4} 2N_5\left( \frac{4}{3}g_3^4 + \frac{1}{15}g_1^4\right) \left|\frac{F}{M}\right|^2 1_{3\times 3} \,,
 \\
 \tilde m^2_{l} &= \frac{1}{(4\pi)^4} 2N_5\left( \frac{3}{4} g_2^4 + \frac{3}{20}g_1^4 \right) \left|\frac{F}{M}\right|^2 1_{3\times 3} \,,
 \\
 \tilde m^2_{e} &= \frac{1}{(4\pi)^4} 2N_5\left( \frac{3}{5}g_1^4\right) \left|\frac{F}{M}\right|^2 1_{3\times 3} \,,
 \end{align}
at the messenger scale, where $x=F/M^2$ and~\cite{Martin:1996zb}
\begin{equation}
 g(x) = \frac{1}{x^2}\left[(1+x)\log(1+x)+(1-x)\log(1-x)\right] = 1 + \frac{x^2}6 + \Order(x^4).
\end{equation}
We did not include the corresponding term for sfermion masses~\cite{Dimopoulos:1996gy} in our numerical calculations.
This is for consistency of the two-loop messenger--matter terms that are leading order in $F/M^2$.

Due to the messenger--matter interactions these parameters are corrected.
The one- and two-loop contributions at the messenger scale are summarized as
\begin{align}\label{app_sqmasses}
 \delta \tilde m^2_{H_U} & = - \frac{3}{(4 \pi)^4} \left[ \Tr \bigl( Y_u^{\dag} y_u y_u^{\dag} Y_u \bigr) +2 \Tr \bigl( Y_u y_u^{\dag} y_u Y_u^{\dag} \bigr) \right] \left|\frac{F}{M}\right|^2 ,
 \\
 \delta \tilde m^2_{H_D} & = - \frac{3}{(4 \pi)^4} \Tr \bigl( Y_d^{\dag} y_u y_u^{\dag} Y_d \bigr) \left|\frac{F}{M}\right|^2 ,
 \\
\begin{split}
  \delta \tilde m^2_q &= -\frac{1}{(4\pi)^2} \frac{1}{6} \bigl( y_u y_u^{\dagger} \bigr) \frac{F^4}{M^6} \, h(x)
 \\&\quad
 +\frac{1}{(4\pi)^4} \Bigg\{
 \left[ 3\Tr\bigl( y_u^{\dagger}y_u\bigr)
 -\frac{16}{3}g_3^2 - 3g_2^2 - \frac{13}{15}g_1^2\right] y_u y_u^{\dagger}
 +3y_u y_u^{\dagger}y_u y_u^{\dagger} 
 +2y_u Y_u^{\dagger} Y_u y_u^{\dagger} 
 \\ &\qquad
 -2Y_u y_u^{\dagger} y_u Y_u^{\dagger}
 +y_u Y_u^{\dagger} \Tr\bigl( 3y_u^{\dagger} Y_u \bigr) 
 +Y_u y_u^{\dagger} \Tr\bigl( 3Y_u^{\dagger}y_u \bigr) 
\Bigg\}
 \left|\frac{F}{M}\right|^2,
\end{split}\\
\begin{split}
 \delta\tilde m^2_{u_R} 
 &=
 -\frac{1}{(4\pi)^2}
 \frac{1}{3} \bigl( y_u^\dagger y_u \bigr) 
 \frac{F^4}{M^6} \, h(x)
 \\&\quad
 +\frac{1}{(4\pi)^4}
 \Bigg\{
 2\left[ 3\Tr\bigl( y_u^{\dagger}y_u \bigr) 
 -\frac{16}{3}g_3^2 - 3g_2^2 - \frac{13}{15}g_1^2\right] y_u^{\dagger}y_u
 +6 y_u^{\dagger}y_u y_u^{\dagger}y_u 
 +2y_u^{\dagger}Y_u Y_u^\dagger y_u 
 \\&\qquad
 +2y_u^{\dagger}Y_d Y_d^\dagger y_u 
 -2Y_u^\dagger y_uy_u^{\dagger}Y_u 
 +2y_u^{\dagger}Y_u \Tr\bigl( 3Y_u^\dagger y_u \bigr) 
 +2Y_u^\dagger y_u \Tr\bigl( 3y_u^{\dagger}Y_u\bigr) 
 \Bigg\}
 \left|\frac{F}{M}\right|^2,
\end{split} 
\\
 \delta\tilde m^2_{d_R}
 &=
 -\frac{1}{(4\pi)^4}
 2Y_d^\dagger y_u y_u^{\dagger}Y_d
 \left|\frac{F}{M}\right|^2 ,
 \\
 \delta m^2_{l}&=\delta m^2_{e_R} = 0.
 \end{align}
 In the expressions for $\delta \tilde m^2_q$ and $\delta \tilde m^2_{u_R}$, the first line is the one-loop contribution suppressed by a factor of $x^2$ over the 2-loop and the GMSB contributions with~\cite{Dubovsky:1997rq}
 \[
 h(x) = - 3\frac{(2-x)\log(1-x) + (2+x)\log(1+x)}{x^4}
      = 1 + \frac{4x^2}{5} + \Order(x^4) .
 \]

\enlargethispage{2\baselineskip}
 In addition, the A-terms receive the one-loop contributions
 \begin{align}
 A^*_u
 &=
 -\frac{1}{16\pi^2}\left[
 \left( y_u y_u^{\dagger} \right) Y_u
 +2Y_u\left( y_u^{\dagger}y_u \right) 
 \right]\frac{F}{M}\,,
 \\
 A^*_d &= 
 -\frac{1}{16\pi^2}\left[
 \left( y_u y_u^{\dagger} \right) Y_d
 \right]\frac{F}{M}\,. 
 \end{align}

 \section{Higgs Mass Validation}
 \label{sec:feynhiggs}
 
Since the computation of the Higgs mass in some regions of the parameter space 
involves various subtleties, we also computed it with 
\feynhiggs~\cite{Heinemeyer:1998yj,Heinemeyer:1998np,Degrassi:2002fi,Frank:2006yh,Hahn:2013ria} for our examples.
The comparison in \tableref{feynhiggs} shows that for the selected points 
in parameter space, the two codes are in good agreement, 
with the largest discrepancies for the heavy 7 TeV 
squarks spectrum of \figref{U22} and the mixed 
$\tilde t - \tilde c$ spectrum of \figref{U32-ct}.
For spectra in the multi-TeV range, this may be due to the implementation of resummation in \feynhiggs.

 \begin{table}[tp]
  \catcode`?=\active \def?{\phantom{0}}
  \centering
 \begin{tabular}{|c@{ }c|cc|}
 \hline
 \multicolumn{2}{|c|}{BP} & \softsusy\ $m_h$ [GeV] & \feynhiggs\ $m_h$ [GeV] 
 \\ \hline
 MFV-t  & (\figref{MFV-t})            & $124.6$ & $ 123.5 \pm 2.6$ \\
 MFV-ct & (\figref{MFV-ct})            & $125.0$ & $ 121.6 \pm 3.6$ \\
 \hline
 U11    & (\figref{U11})      & $118.0$ & $ 117.8 \pm 2.6$ \\
 \hline
 U22L     & (\figref{U22L})      & $117.8$ & $ 117.7 \pm 2.6$ \\
 U22      & (\figref{U22})         & $124.6$ & $ 120.2 \pm 5.5$ \\
 \hline
 U32-c    & (\figref{U32-c}) & $125.0$ & $ 123.3 \pm 3.5$ \\
 U32-t    & (\figref{U32-t})   & $124.9$ & $ 123.3 \pm 3.0$ \\
 U32-ct   & (\figref{U32-ct}) & $125.7$ & $ 120.2 \pm 2.6$ \\
 \hline
 \end{tabular}
 \caption{Comparison of the Higgs mass obtained from \softsusy\ and \feynhiggs.}
 \label{tab:feynhiggs}
 \end{table}

\section{Calculation of Flavor-Violating Parameters} 
\label{sec:app_deltas}

In this work, following \cite{Galon:2013jba}, we have used the formula
\begin{equation}
  \bigl( \delta_{ij}^q \bigr) _{MM} = \frac{\Delta \tilde m^2_{ji}}{\tilde m^2_q} \bigl( K_M^q \bigr) _{ij} \bigl( K_M^q \bigr) ^* _{jj}
\label{deltaformula2}
\end{equation}
to parameterize the flavor violation in the $M$-handed squark sector, where we ignore the LR mixing.
Here $\Delta \tilde m^2_{ji} = m^2_{\tilde q_j} - m^2_{\tilde q_i}$ is the squared-mass difference of $M$-handed squarks, $\tilde m^2_q = \frac{1}{3} \sum_{\alpha=1}^3 m^2_{\tilde q_\alpha}$ is the average mass, and $(K_M^q)_{ij}$ is the mixing appearing in the quark--squark--gluino coupling.
This formula is obtained from the MIA formula~\cite{Hiller:2008sv}
\begin{equation}
 \sum_{a} \frac{\Delta \tilde m^2_{a}}{\tilde m^2_q} \bigl( K_M^q \bigr) _{ia} \bigl( K_M^q \bigr) ^* _{ja} \,,
  \label{deltaformula3}
\end{equation}
where $\Delta \tilde m_a^2=m^2_{\tilde q_a}-\tilde m^2_q$.
In some of our spectra, there are noticeable differences in the values obtained for $\delta$ using the two formulae:
this typically occurs when there are sizable mass splittings between or mixings of the first two generations and the third generation.
Thus, as a reference, we show the values of $(\delta^{u}_{12})_{MM}$ in both of the formulae in \tableref{deltau12}.
The other parameters $(\delta^{d}_{12})_{MM}$ and $(\delta^{d}_{23})_{MM}$ are not shown as they are far below the bounds in \tableref{boundsMM}.
 All the spectra satisfy the bound on $|(\delta^u_{12})|_{MM}$ in \tableref{boundsMM},
 which is calculated for $\tilde m_q=1\TeV$ and scales approximately linearly with $\tilde m_q$.
 When one takes into account this scaling, the BPs U11, U22L, and U22 have values of 
 $\sqrt{|(\delta^u_{12})_{LL}||(\delta^u_{12})_{RR}|}$ that exceed the bounds by factors of 2--3 
 assuming $\mathcal{O}(1)$ phases; therefore, these examples will require mild suppression of 
 their phases \cite{Hiller:2008sv, Isidori:2010kg}. 

\begin{table}[t]
  \catcode`?=\active \def?{\phantom{0}}
 \caption{%
 $(\delta^u_{12})_{MM}$ for the example spectra in \secsref{MFV}{NonMFV} given by the two formulae 
 \eqref{deltaformula2} and \eqref{deltaformula3}.
 }
 \label{tab:deltau12}\centering
 \begin{tabular}[t]{|c@{ }cc|ccc|}
 \hline
 \multicolumn{2}{|c}{BP} & formula 
 & $|(\delta^u_{12})_{LL}|$ & $|(\delta^u_{12})_{RR}|$ & $\sqrt{|(\delta^u_{12})_{LL}||(\delta^u_{12})_{RR}|}$
 \\ \hline
 MFV-t   & (\figref{MFV-t}) & 
   \ref{deltaformula2} & 0.0018  & 0.0063 & 0.0033 \\
 & & \ref{deltaformula3} & 0.0010  & $2.3 \times 10^{-5}$ & $1.6 \times 10^{-4}$ \\\hline
 MFV-ct  & (\figref{MFV-ct}) & 
   \ref{deltaformula2} & $3.8 \times 10^{-4}$ & 0.0077 & 0.0017 \\
 & & \ref{deltaformula3} & $6.0 \times 10^{-4}$ & 0.022?   & 0.0036 \\\hline
 U11     & (\figref{U11})& 
   \ref{deltaformula2} & 0.10  & 0.0011   & 0.011 \\
 & & \ref{deltaformula3} & 0.10  & 0.0011   & 0.011 \\\hline
 U22L  & (\figref{U22L}) & 
   \ref{deltaformula2} & 0.099 & 0.0020   & 0.014 \\
 & & \ref{deltaformula3} & 0.099 & 0.0020   & 0.014 \\\hline
 U22    & (\figref{U22}) & 
   \ref{deltaformula2} & 0.12  & 0.0023   & 0.016 \\
 & & \ref{deltaformula3} & 0.12  & 0.0023   & 0.016 \\\hline
 U32-c  & (\figref{U32-c}) & 
          \ref{deltaformula2} & 0.0040  & 0.0026   & 0.0032   \\
  & & \ref{deltaformula3} & $2.7 \times 10^{-4}$ & 0.0026 & $8.4 \times 10^{-4}$ \\\hline
 U32-t  & (\figref{U32-t}) & 
   \ref{deltaformula2} & $1.9 \times 10^{-4}$ & $7.4 \times 10^{-4}$ & $3.7 \times 10^{-4}$ \\
 & & \ref{deltaformula3} & $5.7 \times 10^{-5}$ & $7.4 \times 10^{-4}$  & $2.0 \times 10^{-4}$ \\\hline 
 U32-ct & (\figref{U32-ct}) & 
   \ref{deltaformula2} & 0.0010  & 0.0045   & 0.0022   \\
 & & \ref{deltaformula3} & $2.8 \times 10^{-4}$ & 0.0024 & $8.1 \times 10^{-4}$ \\\hline
\end{tabular}\end{table}

 \bibliography{bibliography_jhep.bib}

\providecommand{\href}[2]{#2}\begingroup\raggedright\begin{thebibliography}{10}

\bibitem{Shadmi:2011hs}
Y.~Shadmi and P.~Z. Szabo, {\it {Flavored Gauge-Mediation}},  {\em JHEP} {\bf
  06} (2012) 124, [\href{http://arxiv.org/abs/1103.0292}{{\tt
  arXiv:1103.0292}}].

\bibitem{Dine:1994vc}
M.~Dine, A.~E. Nelson, and Y.~Shirman, {\it {Low-energy dynamical supersymmetry
  breaking simplified}},  {\em Phys. Rev.} {\bf D51} (1995) 1362--1370,
  [\href{http://arxiv.org/abs/hep-ph/9408384}{{\tt hep-ph/9408384}}].

\bibitem{Dine:1995ag}
M.~Dine, A.~E. Nelson, Y.~Nir, and Y.~Shirman, {\it {New tools for low-energy
  dynamical supersymmetry breaking}},  {\em Phys. Rev.} {\bf D53} (1996)
  2658--2669, [\href{http://arxiv.org/abs/hep-ph/9507378}{{\tt
  hep-ph/9507378}}].

\bibitem{Dine:1996xk}
M.~Dine, Y.~Nir, and Y.~Shirman, {\it {Variations on minimal gauge mediated
  supersymmetry breaking}},  {\em Phys. Rev.} {\bf D55} (1997) 1501--1508,
  [\href{http://arxiv.org/abs/hep-ph/9607397}{{\tt hep-ph/9607397}}].

\bibitem{Chacko:2001km}
Z.~Chacko and E.~Ponton, {\it {Yukawa deflected gauge mediation}},  {\em Phys.
  Rev.} {\bf D66} (2002) 095004,
  [\href{http://arxiv.org/abs/hep-ph/0112190}{{\tt hep-ph/0112190}}].

\bibitem{Joaquim:2006uz}
F.~R. Joaquim and A.~Rossi, {\it {Gauge and Yukawa mediated supersymmetry
  breaking in the triplet seesaw scenario}},  {\em Phys. Rev. Lett.} {\bf 97}
  (2006) 181801, [\href{http://arxiv.org/abs/hep-ph/0604083}{{\tt
  hep-ph/0604083}}].

\bibitem{Joaquim:2006mn}
F.~R. Joaquim and A.~Rossi, {\it {Phenomenology of the triplet seesaw mechanism
  with Gauge and Yukawa mediation of SUSY breaking}},  {\em Nucl. Phys.} {\bf
  B765} (2007) 71--117, [\href{http://arxiv.org/abs/hep-ph/0607298}{{\tt
  hep-ph/0607298}}].

\bibitem{Brignole:2010nh}
A.~Brignole, F.~R. Joaquim, and A.~Rossi, {\it {Beyond the standard seesaw:
  Neutrino masses from Kahler operators and broken supersymmetry}},  {\em JHEP}
  {\bf 08} (2010) 133, [\href{http://arxiv.org/abs/1007.1942}{{\tt
  arXiv:1007.1942}}].

\bibitem{Evans:2011bea}
J.~L. Evans, M.~Ibe, and T.~T. Yanagida, {\it {Relatively Heavy Higgs Boson in
  More Generic Gauge Mediation}},  {\em Phys. Lett.} {\bf B705} (2011)
  342--348, [\href{http://arxiv.org/abs/1107.3006}{{\tt arXiv:1107.3006}}].

\bibitem{Evans:2011uq}
J.~L. Evans, M.~Ibe, and T.~T. Yanagida, {\it {Probing Extra Matter in Gauge
  Mediation Through the Lightest Higgs Boson Mass}},
  \href{http://arxiv.org/abs/1108.3437}{{\tt arXiv:1108.3437}}.

\bibitem{Evans:2012hg}
J.~L. Evans, M.~Ibe, S.~Shirai, and T.~T. Yanagida, {\it {A 125GeV Higgs Boson
  and Muon g-2 in More Generic Gauge Mediation}},  {\em Phys. Rev.} {\bf D85}
  (2012) 095004, [\href{http://arxiv.org/abs/1201.2611}{{\tt
  arXiv:1201.2611}}].

\bibitem{Kang:2012ra}
Z.~Kang, T.~Li, T.~Liu, C.~Tong, and J.~M. Yang, {\it {A Heavy SM-like Higgs
  and a Light Stop from Yukawa-Deflected Gauge Mediation}},  {\em Phys. Rev.}
  {\bf D86} (2012) 095020, [\href{http://arxiv.org/abs/1203.2336}{{\tt
  arXiv:1203.2336}}].

\bibitem{Evans:2012uf}
J.~L. Evans, M.~Ibe, and T.~T. Yanagida, {\it {The Lightest Higgs Boson Mass in
  the MSSM with Strongly Interacting Spectators}},  {\em Phys. Rev.} {\bf D86}
  (2012) 015017, [\href{http://arxiv.org/abs/1204.6085}{{\tt
  arXiv:1204.6085}}].

\bibitem{Craig:2012xp}
N.~Craig, S.~Knapen, D.~Shih, and Y.~Zhao, {\it {A Complete Model of Low-Scale
  Gauge Mediation}},  {\em JHEP} {\bf 03} (2013) 154,
  [\href{http://arxiv.org/abs/1206.4086}{{\tt arXiv:1206.4086}}].

\bibitem{Albaid:2012qk}
A.~Albaid and K.~S. Babu, {\it {Higgs boson of mass 125 GeV in GMSB models with
  messenger-matter mixing}},  {\em Phys. Rev.} {\bf D88} (2013) 055007,
  [\href{http://arxiv.org/abs/1207.1014}{{\tt arXiv:1207.1014}}].

\bibitem{Abdullah:2012tq}
M.~Abdullah, I.~Galon, Y.~Shadmi, and Y.~Shirman, {\it {Flavored Gauge
  Mediation, A Heavy Higgs, and Supersymmetric Alignment}},  {\em JHEP} {\bf
  06} (2013) 057, [\href{http://arxiv.org/abs/1209.4904}{{\tt
  arXiv:1209.4904}}].

\bibitem{Byakti:2013ti}
P.~Byakti and T.~S. Ray, {\it {Burgeoning the Higgs mass to 125 GeV through
  messenger-matter interactions in GMSB models}},  {\em JHEP} {\bf 05} (2013)
  055, [\href{http://arxiv.org/abs/1301.7605}{{\tt arXiv:1301.7605}}].

\bibitem{Craig:2013wga}
N.~Craig, S.~Knapen, and D.~Shih, {\it {General Messenger Higgs Mediation}},
  {\em JHEP} {\bf 08} (2013) 118, [\href{http://arxiv.org/abs/1302.2642}{{\tt
  arXiv:1302.2642}}].

\bibitem{Evans:2013kxa}
J.~A. Evans and D.~Shih, {\it {Surveying Extended GMSB Models with
  $m$$_{h}$=125 GeV}},  {\em JHEP} {\bf 08} (2013) 093,
  [\href{http://arxiv.org/abs/1303.0228}{{\tt arXiv:1303.0228}}].

\bibitem{Jelinski:2015voa}
T.~Jelinski and J.~Gluza, {\it {Analytical two-loop soft mass terms of
  sfermions in Extended GMSB models}},  {\em Phys. Lett.} {\bf B751} (2015)
  541--547, [\href{http://arxiv.org/abs/1505.07443}{{\tt arXiv:1505.07443}}].

\bibitem{Basirnia:2015vga}
A.~Basirnia, D.~Egana-Ugrinovic, S.~Knapen, and D.~Shih, {\it {125 GeV Higgs
  from Tree-Level $A$-terms}},  {\em JHEP} {\bf 06} (2015) 144,
  [\href{http://arxiv.org/abs/1501.00997}{{\tt arXiv:1501.00997}}].

\bibitem{Calibbi:2013mka}
L.~Calibbi, P.~Paradisi, and R.~Ziegler, {\it {Gauge Mediation beyond Minimal
  Flavor Violation}},  {\em JHEP} {\bf 06} (2013) 052,
  [\href{http://arxiv.org/abs/1304.1453}{{\tt arXiv:1304.1453}}].

\bibitem{Calibbi:2014yha}
L.~Calibbi, P.~Paradisi, and R.~Ziegler, {\it {Lepton Flavor Violation in
  Flavored Gauge Mediation}},  {\em Eur. Phys. J.} {\bf C74} (2014), no.~12
  3211, [\href{http://arxiv.org/abs/1408.0754}{{\tt arXiv:1408.0754}}].

\bibitem{Galon:2013jba}
I.~Galon, G.~Perez, and Y.~Shadmi, {\it {Non-Degenerate Squarks from Flavored
  Gauge Mediation}},  {\em JHEP} {\bf 1309} (2013) 117,
  [\href{http://arxiv.org/abs/1306.6631}{{\tt arXiv:1306.6631}}].

\bibitem{Mahbubani:2012qq}
R.~Mahbubani, M.~Papucci, G.~Perez, J.~T. Ruderman, and A.~Weiler, {\it {Light
  Nondegenerate Squarks at the LHC}},  {\em Phys. Rev. Lett.} {\bf 110} (2013),
  no.~15 151804, [\href{http://arxiv.org/abs/1212.3328}{{\tt
  arXiv:1212.3328}}].

\bibitem{Blanke:2013zxo}
M.~Blanke, G.~F. Giudice, P.~Paradisi, G.~Perez, and J.~Zupan, {\it {Flavoured
  Naturalness}},  {\em JHEP} {\bf 06} (2013) 022,
  [\href{http://arxiv.org/abs/1302.7232}{{\tt arXiv:1302.7232}}].

\bibitem{Blanke:2015ulx}
M.~Blanke, B.~Fuks, I.~Galon, and G.~Perez, {\it {Gluino Meets Flavored
  Naturalness}},  \href{http://arxiv.org/abs/1512.03813}{{\tt
  arXiv:1512.03813}}.

\bibitem{Brummer:2013upa}
F.~Brümmer, M.~McGarrie, and A.~Weiler, {\it {Light third-generation squarks
  from flavour gauge messengers}},  {\em JHEP} {\bf 04} (2014) 078,
  [\href{http://arxiv.org/abs/1312.0935}{{\tt arXiv:1312.0935}}].

\bibitem{Jelinski:2014uba}
T.~Jelinski and J.~Pawelczyk, {\it {Masses and FCNC in Flavoured GMSB scheme}},
   \href{http://arxiv.org/abs/1406.4001}{{\tt arXiv:1406.4001}}.

\bibitem{Abel:2014fka}
S.~Abel and M.~McGarrie, {\it {Natural supersymmetry and dynamical flavour with
  meta-stable vacua}},  {\em JHEP} {\bf 07} (2014) 145,
  [\href{http://arxiv.org/abs/1404.1318}{{\tt arXiv:1404.1318}}].

\bibitem{Evans:2015swa}
J.~A. Evans, D.~Shih, and A.~Thalapillil, {\it {Chiral Flavor Violation from
  Extended Gauge Mediation}},  {\em JHEP} {\bf 07} (2015) 040,
  [\href{http://arxiv.org/abs/1504.00930}{{\tt arXiv:1504.00930}}].

\bibitem{Backovic:2015rwa}
M.~Backović, A.~Mariotti, and M.~Spannowsky, {\it {Signs of Tops from Highly
  Mixed Stops}},  {\em JHEP} {\bf 06} (2015) 122,
  [\href{http://arxiv.org/abs/1504.00927}{{\tt arXiv:1504.00927}}].

\bibitem{Feng:2007ke}
J.~L. Feng, C.~G. Lester, Y.~Nir, and Y.~Shadmi, {\it {The Standard Model and
  Supersymmetric Flavor Puzzles at the Large Hadron Collider}},  {\em Phys.
  Rev.} {\bf D77} (2008) 076002, [\href{http://arxiv.org/abs/0712.0674}{{\tt
  arXiv:0712.0674}}].

\bibitem{Nir:1993mx}
Y.~Nir and N.~Seiberg, {\it {Should squarks be degenerate?}},  {\em Phys.
  Lett.} {\bf B309} (1993) 337--343,
  [\href{http://arxiv.org/abs/hep-ph/9304307}{{\tt hep-ph/9304307}}].

\bibitem{Ghosh:2015nui}
D.~Ghosh, P.~Paradisi, G.~Perez, and G.~Spada, {\it {CP Violation Tests of
  Alignment Models at LHCII}},  \href{http://arxiv.org/abs/1512.03962}{{\tt
  arXiv:1512.03962}}.

\bibitem{Evans:2016zau}
J.~A. Evans and J.~Shelton, {\it {Long-Lived Staus and Displaced Leptons at the
  LHC}},  \href{http://arxiv.org/abs/1601.01326}{{\tt arXiv:1601.01326}}.

\bibitem{Allanach:2001kg}
B.~C. Allanach, {\it {SOFTSUSY: a program for calculating supersymmetric
  spectra}},  {\em Comput. Phys. Commun.} {\bf 143} (2002) 305--331,
  [\href{http://arxiv.org/abs/hep-ph/0104145}{{\tt hep-ph/0104145}}].

\bibitem{Heinemeyer:1998yj}
S.~Heinemeyer, W.~Hollik, and G.~Weiglein, {\it {FeynHiggs: A Program for the
  calculation of the masses of the neutral CP even Higgs bosons in the MSSM}},
  {\em Comput. Phys. Commun.} {\bf 124} (2000) 76--89,
  [\href{http://arxiv.org/abs/hep-ph/9812320}{{\tt hep-ph/9812320}}].

\bibitem{Heinemeyer:1998np}
S.~Heinemeyer, W.~Hollik, and G.~Weiglein, {\it {The Masses of the neutral CP -
  even Higgs bosons in the MSSM: Accurate analysis at the two loop level}},
  {\em Eur. Phys. J.} {\bf C9} (1999) 343--366,
  [\href{http://arxiv.org/abs/hep-ph/9812472}{{\tt hep-ph/9812472}}].

\bibitem{Degrassi:2002fi}
G.~Degrassi, S.~Heinemeyer, W.~Hollik, P.~Slavich, and G.~Weiglein, {\it
  {Towards high precision predictions for the MSSM Higgs sector}},  {\em Eur.
  Phys. J.} {\bf C28} (2003) 133--143,
  [\href{http://arxiv.org/abs/hep-ph/0212020}{{\tt hep-ph/0212020}}].

\bibitem{Frank:2006yh}
M.~Frank, T.~Hahn, S.~Heinemeyer, W.~Hollik, H.~Rzehak, and G.~Weiglein, {\it
  {The Higgs Boson Masses and Mixings of the Complex MSSM in the
  Feynman-Diagrammatic Approach}},  {\em JHEP} {\bf 02} (2007) 047,
  [\href{http://arxiv.org/abs/hep-ph/0611326}{{\tt hep-ph/0611326}}].

\bibitem{Hahn:2013ria}
T.~Hahn, S.~Heinemeyer, W.~Hollik, H.~Rzehak, and G.~Weiglein, {\it
  {High-Precision Predictions for the Light CP -Even Higgs Boson Mass of the
  Minimal Supersymmetric Standard Model}},  {\em Phys. Rev. Lett.} {\bf 112}
  (2014), no.~14 141801, [\href{http://arxiv.org/abs/1312.4937}{{\tt
  arXiv:1312.4937}}].

\bibitem{Cao:2006xb}
J.~Cao, G.~Eilam, K.-i. Hikasa, and J.~M. Yang, {\it {Experimental constraints
  on stop-scharm flavor mixing and implications in top-quark FCNC processes}},
  {\em Phys. Rev.} {\bf D74} (2006) 031701,
  [\href{http://arxiv.org/abs/hep-ph/0604163}{{\tt hep-ph/0604163}}].

\bibitem{Kowalska:2014opa}
K.~Kowalska, {\it {Phenomenology of SUSY with General Flavour Violation}},
  {\em JHEP} {\bf 09} (2014) 139, [\href{http://arxiv.org/abs/1406.0710}{{\tt
  arXiv:1406.0710}}].

\bibitem{Brignole:2015kva}
A.~Brignole, {\it {The supersymmetric Higgs boson with flavoured A-terms}},
  {\em Nucl. Phys.} {\bf B898} (2015) 644--658,
  [\href{http://arxiv.org/abs/1504.03273}{{\tt arXiv:1504.03273}}].

\bibitem{Goodsell:2015yca}
M.~D. Goodsell, K.~Nickel, and F.~Staub, {\it {The Higgs Mass in the MSSM at
  two-loop order beyond minimal flavour violation}},
  \href{http://arxiv.org/abs/1511.01904}{{\tt arXiv:1511.01904}}.

\bibitem{Isidori:2010kg}
G.~Isidori, Y.~Nir, and G.~Perez, {\it {Flavor Physics Constraints for Physics
  Beyond the Standard Model}},  {\em Ann. Rev. Nucl. Part. Sci.} {\bf 60}
  (2010) 355, [\href{http://arxiv.org/abs/1002.0900}{{\tt arXiv:1002.0900}}].

\bibitem{Raz:2002zx}
G.~Raz, {\it {The Mass insertion approximation without squark degeneracy}},
  {\em Phys. Rev.} {\bf D66} (2002) 037701,
  [\href{http://arxiv.org/abs/hep-ph/0205310}{{\tt hep-ph/0205310}}].

\bibitem{Calibbi:2015kja}
L.~Calibbi, I.~Galon, A.~Masiero, P.~Paradisi, and Y.~Shadmi, {\it {Charged
  Slepton Flavor post the 8 TeV LHC: A Simplified Model Analysis of Low-Energy
  Constraints and LHC SUSY Searches}},  {\em JHEP} {\bf 10} (2015) 043,
  [\href{http://arxiv.org/abs/1502.07753}{{\tt arXiv:1502.07753}}].

\bibitem{ATL-PHYS-PUB-2014-010}
{\it {Search for Supersymmetry at the high luminosity LHC with the ATLAS
  experiment}},  Tech. Rep. ATL-PHYS-PUB-2014-010, CERN, Geneva, Jul, 2014.

\bibitem{CMS-PAS-SUS-14-012}
{\bf CMS Collaboration} Collaboration, {\it {Supersymmetry discovery potential
  in future LHC and HL-LHC running with the CMS detector}},  Tech. Rep.
  CMS-PAS-SUS-14-012, CERN, Geneva, 2015.

\bibitem{Aad:2015pfx}
{\bf ATLAS} Collaboration, G.~Aad et~al., {\it {ATLAS Run 1 searches for direct
  pair production of third-generation squarks at the Large Hadron Collider}},
  {\em Eur. Phys. J.} {\bf C75} (2015), no.~10 510,
  [\href{http://arxiv.org/abs/1506.08616}{{\tt arXiv:1506.08616}}].

\bibitem{Khachatryan:2016pup}
{\bf CMS} Collaboration, V.~Khachatryan et~al., {\it {Search for direct pair
  production of scalar top quarks in the single- and dilepton channels in
  proton-proton collisions at sqrt(s) = 8 TeV}},
  \href{http://arxiv.org/abs/1602.03169}{{\tt arXiv:1602.03169}}.

\bibitem{Aad:2016jxj}
{\bf ATLAS} Collaboration, G.~Aad et~al., {\it {Search for new phenomena in
  final states with large jet multiplicities and missing transverse momentum
  with ATLAS using $\sqrt{s} =13$ TeV proton--proton collisions}},
  \href{http://arxiv.org/abs/1602.06194}{{\tt arXiv:1602.06194}}.

\bibitem{Khachatryan:2016kdk}
{\bf CMS} Collaboration, V.~Khachatryan et~al., {\it {Search for supersymmetry
  in the multijet and missing transverse momentum final state in pp collisions
  at 13 TeV}},  \href{http://arxiv.org/abs/1602.06581}{{\tt arXiv:1602.06581}}.

\bibitem{Aad:2015hea}
{\bf ATLAS} Collaboration, G.~Aad et~al., {\it {Search for photonic signatures
  of gauge-mediated supersymmetry in 8 TeV pp collisions with the ATLAS
  detector}},  {\em Phys. Rev.} {\bf D92} (2015), no.~7 072001,
  [\href{http://arxiv.org/abs/1507.05493}{{\tt arXiv:1507.05493}}].

\bibitem{Khachatryan:2016hns}
{\bf CMS} Collaboration, V.~Khachatryan et~al., {\it {Search for supersymmetry
  in electroweak production with photons and large missing transverse energy in
  pp collisions at $\sqrt{s}$ = 8 TeV}},
  \href{http://arxiv.org/abs/1602.08772}{{\tt arXiv:1602.08772}}.

\bibitem{Nir:2002ah}
Y.~Nir and G.~Raz, {\it {Quark squark alignment revisited}},  {\em Phys. Rev.}
  {\bf D66} (2002) 035007, [\href{http://arxiv.org/abs/hep-ph/0206064}{{\tt
  hep-ph/0206064}}].

\bibitem{ATLAS:2014fka}
{\bf ATLAS} Collaboration, {\it {Searches for heavy long-lived charged
  particles with the ATLAS detector in proton-proton collisions at $ \sqrt{s}=8
  $ TeV}},  {\em JHEP} {\bf 01} (2015) 068,
  [\href{http://arxiv.org/abs/1411.6795}{{\tt arXiv:1411.6795}}].

\bibitem{Chatrchyan:2012jwg}
{\bf CMS} Collaboration, S.~Chatrchyan et~al., {\it {Search for long-lived
  particles decaying to photons and missing energy in proton-proton collisions
  at $\sqrt{s}=7$ TeV}},  {\em Phys. Lett.} {\bf B722} (2013) 273--294,
  [\href{http://arxiv.org/abs/1212.1838}{{\tt arXiv:1212.1838}}].

\bibitem{Aad:2014vma}
{\bf ATLAS} Collaboration, {\it {Search for direct production of charginos,
  neutralinos and sleptons in final states with two leptons and missing
  transverse momentum in $pp$ collisions at $\sqrt{s} =$ 8 TeV with the ATLAS
  detector}},  {\em JHEP} {\bf 05} (2014) 071,
  [\href{http://arxiv.org/abs/1403.5294}{{\tt arXiv:1403.5294}}].

\bibitem{Khachatryan:2014qwa}
{\bf CMS} Collaboration, {\it {Searches for electroweak production of
  charginos, neutralinos, and sleptons decaying to leptons and W, Z, and Higgs
  bosons in pp collisions at 8 TeV}},  {\em Eur. Phys. J.} {\bf C74} (2014),
  no.~9 3036, [\href{http://arxiv.org/abs/1405.7570}{{\tt arXiv:1405.7570}}].

\bibitem{Aad:2015wqa}
{\bf ATLAS} Collaboration, G.~Aad et~al., {\it {Search for supersymmetry in
  events containing a same-flavour opposite-sign dilepton pair, jets, and large
  missing transverse momentum in $\sqrt{s}=8$ TeV pp collisions with the ATLAS
  detector}},  {\em Eur. Phys. J.} {\bf C75} (2015), no.~7 318,
  [\href{http://arxiv.org/abs/1503.03290}{{\tt arXiv:1503.03290}}]. [Erratum:
  Eur. Phys. J.C75,no.10,463(2015)].

\bibitem{Aad:2015gna}
{\bf ATLAS} Collaboration, G.~Aad et~al., {\it {Search for Scalar Charm Quark
  Pair Production in $pp$ Collisions at $\sqrt{s}=$ 8??TeV with the ATLAS
  Detector}},  {\em Phys. Rev. Lett.} {\bf 114} (2015), no.~16 161801,
  [\href{http://arxiv.org/abs/1501.01325}{{\tt arXiv:1501.01325}}].

\bibitem{Beringer:1900zz}
{\bf Particle Data Group} Collaboration, J.~Beringer et~al., {\it {Review of
  Particle Physics (RPP)}},  {\em Phys. Rev.} {\bf D86} (2012) 010001.

\bibitem{Dine:1993yw}
M.~Dine and A.~E. Nelson, {\it {Dynamical supersymmetry breaking at
  low-energies}},  {\em Phys. Rev.} {\bf D48} (1993) 1277--1287,
  [\href{http://arxiv.org/abs/hep-ph/9303230}{{\tt hep-ph/9303230}}].

\bibitem{Martin:1996zb}
S.~P. Martin, {\it {Generalized messengers of supersymmetry breaking and the
  sparticle mass spectrum}},  {\em Phys. Rev.} {\bf D55} (1997) 3177--3187,
  [\href{http://arxiv.org/abs/hep-ph/9608224}{{\tt hep-ph/9608224}}].

\bibitem{Dimopoulos:1996gy}
S.~Dimopoulos, G.~F. Giudice, and A.~Pomarol, {\it {Dark matter in theories of
  gauge mediated supersymmetry breaking}},  {\em Phys. Lett.} {\bf B389} (1996)
  37--42, [\href{http://arxiv.org/abs/hep-ph/9607225}{{\tt hep-ph/9607225}}].

\bibitem{Dubovsky:1997rq}
S.~L. Dubovsky and D.~S. Gorbunov, {\it {Messenger matter mixing and lepton
  flavor violation}},  {\em Phys. Lett.} {\bf B419} (1998) 223--232,
  [\href{http://arxiv.org/abs/hep-ph/9706272}{{\tt hep-ph/9706272}}].

\bibitem{Hiller:2008sv}
G.~Hiller, Y.~Hochberg, and Y.~Nir, {\it {Flavor Changing Processes in
  Supersymmetric Models with Hybrid Gauge- and Gravity-Mediation}},  {\em JHEP}
  {\bf 03} (2009) 115, [\href{http://arxiv.org/abs/0812.0511}{{\tt
  arXiv:0812.0511}}].

\end{thebibliography}\endgroup

\end{document}